\documentclass[preprint,12pt,authoryear]{elsarticle}
\makeatletter
\def\ps@pprintTitle{%
  \let\@oddhead\@empty
  \let\@evenhead\@empty
  \let\@oddfoot\@empty
  \let\@evenfoot\@oddfoot
}
\makeatother

\usepackage{amssymb}
\usepackage{amsmath}
\usepackage[colorlinks=true,linkcolor=blue, pdfauthor=author]{hyperref}
\usepackage{caption}
\usepackage{subcaption}
\usepackage{enumitem}
\usepackage{nomencl}
\usepackage{comment}
\usepackage{setspace}
\usepackage{svg}
\usepackage{soul}
\makenomenclature
\usepackage{etoolbox}
\renewcommand\nomgroup[1]{%
  \item[\bfseries
  \ifstrequal{#1}{G}{Greek symbols}{%
  \ifstrequal{#1}{S}{Subscripts}{%
  \ifstrequal{#1}{P}{Superscripts}{}}}%
]}
\usepackage{graphicx}
\graphicspath{{figures/}}
\journal{}

\newcommand{\posvecbase}{r}
\newcommand{\posveccomp}[1]{\posvecbase_{#1}}
\newcommand{\posvec}{\pmb{\posvecbase}}
\newcommand{\posvecheater}{\pmb{\eta}}
\newcommand{\states}{\text{x}}
\newcommand{\posvecheatbase}{\eta}
\newcommand{\posvecheatcomp}[1]{\posvecheatbase_{#1}}
\newcommand{\observations}{\text{y}}

\begin{document}

\begin{frontmatter}



\title{Bayesian Inference for Estimating Heat Sources through Temperature Assimilation}

\author[1]{Hanieh Mousavi}
%\affiliation[1]{organization={Department of Mechanical and Aerospace Engineering, University of California, Los Angeles},
             %city={Los Angeles},
             %postcode={90095-1597},
             %state={CA},
             %country={United States}}
\author[1]{Jeff D. Eldredge\corref{cor1}}
\ead{jdeldre@ucla.edu}
\affiliation[1]{organization={Mechanical and Aerospace Engineering, University of California, Los Angeles},
             city={Los Angeles},
             postcode={90095-1597},
             state={CA},
             country={USA}}
\cortext[cor1]{Corresponding author.}

\begin{abstract}
This paper introduces a Bayesian inference framework for two-dimensional steady-state heat conduction, focusing on the estimation of unknown distributed heat sources in a thermally-conducting medium with uniform conductivity. The goal is to infer heater locations, strengths, and shapes using temperature assimilation in the Euclidean space, employing a Fourier series to represent each heater's shape. The Markov Chain Monte Carlo (MCMC) method, incorporating the random-walk Metropolis-Hasting algorithm and parallel tempering, is utilized for posterior distribution exploration in both unbounded and wall-bounded domains. Strong correlations between heat strength and heater area prompt caution against simultaneously estimating these two quantities. It is found that multiple solutions arise in cases where the number of temperature sensors is less than the number of unknown states. Moreover, smaller heaters introduce greater uncertainty in estimated strength. The diffusive nature of heat conduction smooths out any deformations in the temperature contours, especially in the presence of multiple heaters positioned near each other, impacting convergence. In wall-bounded domains with Neumann boundary conditions, the inference of heater parameters tends to be more accurate than in unbounded domains.
\end{abstract}


\begin{comment}
\begin{highlights}
\item Bubble growth in superheated seawater with time-varying pressure is studied
\item Non-uniform distribution of salt is considered in the boundary layer via solving the equation of mass concentration
\item Salinity reduces growth rate and degrades thermal performance
\item Presence of salt in seawater elongates the \emph{inertia-controlled} regime
\item Depressurization rate strongly affects growth characteristics
\end{highlights}
\end{comment}

\begin{comment}
\begin{keyword}
Bubble growth \sep Infinite seawater \sep Salt concentration \sep Time-varying pressure \sep Depressurization rate


\end{keyword}
\end{comment}

\end{frontmatter}


\section{Introduction} \label{introduction}

%\begin{equation}
    %\posvec, \posveccomp{i}
%\end{equation}
Inverse problems are a class of mathematical problems that estimate the underlying state of a system based on measurements, leveraging a mathematical model termed observation operator. Based on the notion of well-posed problems articulated by \cite{hadamard1923lectures}, inverse problems possess a major difficulty, namely, they are typically ill-posed: uniqueness, stability, and existence of their solution cannot be guaranteed. Even in the case of the existence of a unique solution, the problem can be unstable under the presence of noise in the measurements. Over the years, a spectrum of techniques has been developed to tackle these intricate inverse problems including the least square methods \citep{ritchie1996current}, Kalman filtering \citep{bezruchko2010extracting}, linear quadratic estimation, shooting algorithm \citep{peifer2007parameter}, Newton method \citep{choi2015interpretation}, regularization methods, and Artificial Neural Network \citep{khan2018quantification}. A comprehensive overview of these methodologies can be found in the work of \cite{calvetti2018inverse}.

In the realm of thermal applications, the inverse heat conduction problem encompasses diverse objectives including the determination of thermal boundary conditions that align with observed temperature history \citep{lei2022inverse}, or reconstruction of the initial temperature profile given the temperature history of the substance \citep{castro2010source}. Another central facet of thermal inverse problems involves estimating heat sources responsible for the observed temperature distribution. It has applications in manufacturing, environmental studies, electronic cooling, and aerospace engineering where the identification of unknown heat sources holds critical significance. Over the past decades, researchers have endeavored to address heat source estimation through different techniques in transient heat conduction scenarios. These methodologies involve a range of techniques such as the conjugate gradient method \citep{su2001heat, ma2012identification} and the extended Kalman filter (EKF) together with the weighted recursive least-squares estimator \citep{chen2007using}. Additionally, methods like the modified Newton–Raphson \citep{lin2007estimation}, the Function Specification Algorithm \citep{janicki2008real}, the Particle Swarm Optimization (PSO) technique \citep{bangian2018optimization}, and Neural Networks \citep{kitano2022constructing} have been applied in the field of heat source estimation.

It is essential to recognize that not all the deterministic techniques previously discussed can address the inherent complexities of inverse problems. An alternative and highly effective probabilistic approach, Bayesian inference, offers a flexible and intuitive way to incorporate prior knowledge into the analysis, effectively complementing noisy data. This is achieved by treating the unknown quantities as random variables and providing a statistical representation of the solution that elucidates a range of possible parameter values. Furthermore, Bayesian inference excels in handling complex models and robustly estimates parameters even with limited data. These abilities make Bayesian analysis a powerful tool in a wide range of applications including but not limited to machine learning \citep{bishop2006pattern}, decision making \citep{berger2013statistical}, bioinformatics \citep{durbin1998biological}, environmental modeling, medical diagnosis, and many other disciplines.

In the context of heat conduction problems, Bayesian inference has been employed with notable success. For instance, \cite{wang2004bayesian}, \cite{wang2004hierarchical}, and \cite{khatoon2023fast} implemented Bayesian inference to address transient heat conduction problem to estimate the heat flux at a boundary. The steady-state heat conduction equation has been tackled using Bayesian inference by \cite{jin2008bayesian} and \cite{cao2022bayesian}, with the latter demonstrating its superiority in terms of accuracy, stability, and robustness compared to traditional deterministic curve-fitting approaches.

This work employs Bayesian inference to tackle the problem of solving the steady-state heat conduction equation in the presence of heat sources within the domain. Unlike previous studies focused on estimating a single-point heat source, our primary objective is to optimally estimate the location, intensity (or strength), size, and shape of one or more regions in which heat is uniformly distributed. Temperature measurements are acquired through the same observation operator at prescribed sensor locations. To effectively represent the heaters' shape and size, a Fourier series is implemented as an approximation. The core optimization technique adopted in this study is the Markov Chain Monte Carlo (MCMC) method, which facilitates sampling from the posterior distribution. Specifically, we utilize a random-walk Metropolis-Hasting (MH) algorithm to traverse the state space. To explore the state space more efficiently, parallel tempering is employed to enable potential exchange between chains. The approach presented here promises to yield valuable insight into applications such as temperature control and optimization in engineering and environmental systems.

In Section \ref{problemstatement}, 
 the problem is presented and the methodology to the Bayesian inference approach to tackle the heat conduction problem is elaborated. The outcomes of the estimation are addressed in Section \ref{results}. Finally, the overal approach and the notable results are summarized in Section \ref{conclusion}.

\section{Problem Statement and methodology}\label{problemstatement}
\begin{figure}[ht!]
\begin{subfigure}{0.3\textwidth}
    \centering
    \includegraphics[width=1\linewidth]{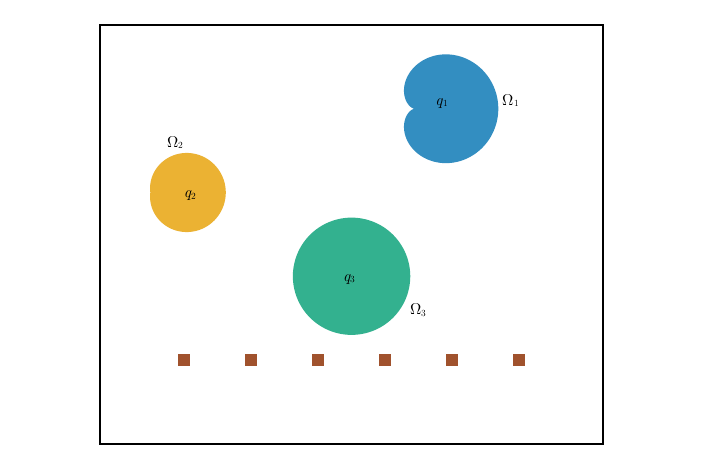} 
    \caption{}
    \label{fig:}
\end{subfigure}
\begin{subfigure}{0.3\textwidth}
    \centering
    \includegraphics[width=1\linewidth]{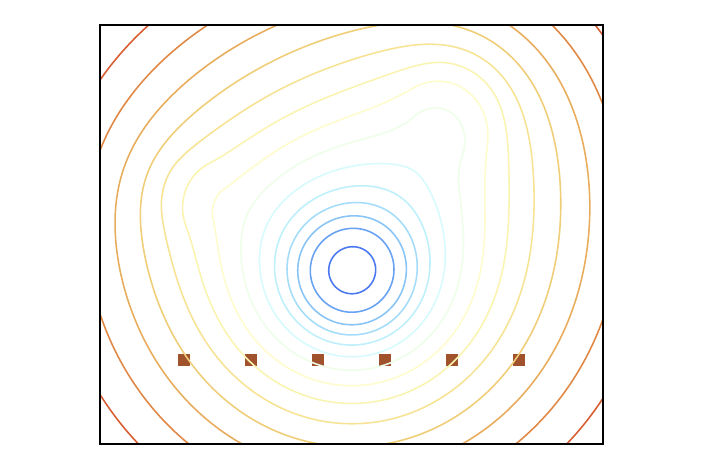} 
    \caption{}
    \label{fig:}
\end{subfigure}
\begin{subfigure}{0.3\textwidth}
    \centering
    \includegraphics[width=1\linewidth]{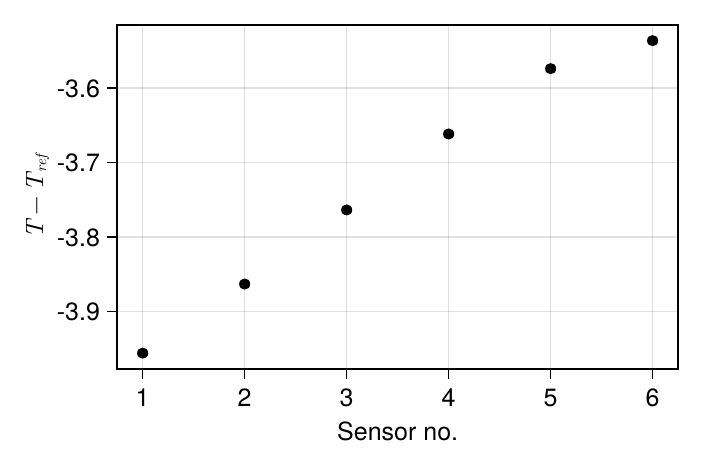} 
    \caption{}
    \label{fig:}
\end{subfigure}

    \caption{\centering Illustration of the problem configuration, featuring three distinct heaters situated within various regions of the 2D domain. Each heater exhibits unique strengths and shapes. The evaluation points are indicated using brown squares, where they can either be placed in an unbounded domain or aligned on an adiabatic wall. (a) represents the heaters in tandem with temperature sensors, (b) temperature contours induced from the corresponding heater configuration in (a), and (c) sensor measurements.}
    \label{fig:config}
\end{figure}

The problem configuration and the mathematical model are addressed in this section. The core objective is to determine the causal heating factors behind steady-state temperature observations in a thermally conducting medium with uniform conductivity, effectively resolving the inverse problem. This problem is characterized by multiple heating regions, $\Omega$, located within a two-dimensional domain depicted in Figure \ref{fig:config}(a). These heaters possess unknown variables encompassing their locations, strengths, sizes, and shapes, resulting in observable steady temperatures at a finite number of designated measurement points. It is important to note that heat is uniformly distributed within each heater. The true measurements are obtained from the same observation operator as the numerical solution of the steady-state heat conduction problem, thereby making it a twin experiment.

In addition to addressing the inverse problem in an unbounded domain, this study investigates the impact of sensors aligned on an adiabatic wall, where the homogeneous Neumann boundary condition applies, on the performance of the inference process. We seek to find the heaters and their corresponding unknown characteristics assimilating temperature observations at the measurement points.

The mathematical model comprises the solution of the steady-state heat conduction equation defined in the Euclidean space
\begin{equation} \label{eq heat}
    \nabla^2 T = - q.
\end{equation}
Here, $T$ denotes temperature and $q$ represents heat source scaled by the thermal conductivity of the underlying material, referred to in this paper as heater \emph{intensity} (or \emph{strength}). The unknowns intrinsic to the problem—--including locations of the center of the heaters as well as their strengths, sizes, and shapes—--are encompassed within the function $q$ in the form of
\begin{equation}
    q(\posvec) = 
    \begin{cases}
        q_h & \quad \posvec \in \Omega_h \\
        0 & \quad \text{otherwise}
    \end{cases},
\end{equation}
where $\posvec=(x,y)$ denotes the position vector in the Euclidean space, and the index refers to the $h^{th}$ heater. The heater shape, $\Omega_h$, is represented by a Fourier series: by dividing the range $0$ to $2 \pi$ into $N$ equally spaced sectors, each with angle $\theta_j$, the coordinates of the respective $N$ points on the boundary of a heater can be expressed in complex notation as:
\begin{equation} \label{eq shape}
    z = x_0 + i y_0 + \sum_{j=1}^\infty c_j e^{i j \theta_j}.
\end{equation}
In this equation, $i$ denotes the imaginary unit, while $c_j$ signify the coefficients of the Fourier series which are assumed to be real numbers in this work. Notably, $x_0+iy_0$ corresponds to the center of the heater shape, with $x_0$ and $y_0$ denoting its position along the two axes within the two-dimensional domain. In this paper, only the first two coefficients are considered and as such, the state vector for the problem of estimating one heater becomes $\states = (x_0, y_0, q_h, c_1, c_2)$ with observation represented as $\observations = (y_1, y_2, \cdots, y_d)$, where $d$ denotes the number of measurement points. Thus, the total number of states when there are $N_q$ heaters in the space will be $n=5N_q$.

\subsection{Observation Operator}
The core component of the estimation procedure is the observation operator, mapping the states to the measurements. In this work, this operator arises from the solution of equation (\ref{eq heat}), the Poisson equation. This equation is solved here on a Cartesian grid using the lattice Green's function, as extensively discussed in works such as \cite{liska2014parallel} and \cite{eldredge2022method}. For the purposes of demonstration in this work, both the synthetic generation of true sensor measurements and the estimation procedure itself rely on the same numerical solution procedure. In scenarios involving boundaries, the Poisson problem is solved using the immersed boundary projection method of \cite{taira2007immersed}, in conjunction with a masking function to distinguish between the two sides of the boundaries. The concept of the masked function in the context of the immersed boundary method was introduced by \cite{eldredge2022method}. The reader is referred to the latter paper for a comprehensive understanding of the solution procedures and operators.

\subsection{Statistical Notation}
We can generalize the solution of Eq. (\ref{eq heat}) to represent a stochastic observation model, given as
\begin{equation} \label{eq forwardModel}
    \observations = h(\states) + \varepsilon ,
\end{equation}
where $\varepsilon$ is the observation noise, and $\states$ is the state vector of unknowns. The operator $h(\states) \in \mathbb{R}^{n \rightarrow d}$ embodies the observation operator establishing a connection between the causes $\states \in \mathbb{R}^n$ and the observations $\observations \in \mathbb{R}^d$, where $n$ denotes the number of unknowns and $d$ represents the number of measurements. As discussed earlier, we employ the Bayesian inference approach to find the state variables, effectively incorporating their uncertainties into the representation.

Drawing upon Bayes' theorem and the definition of a joint probability distribution, the conditional probability of state $\states$, referred to as the \emph{posterior distribution}, can be expressed as follows:
\begin{equation} \label{eq Bayes'}
    \pi(\states|\observations) = \frac{L(\observations|\states) \pi_0(\states)}{\pi(\observations)}.
\end{equation}
This equation establishes a connection for the posterior distribution by considering the probability of $\observations$ conditioned on $\states$ known as \emph{likelihood}, $L(\observations|\states)$, along with the prior probability density of $\states$, $\pi_0(\states)$, and the marginal probability density of the measurements $\observations$. This marginal probability density, denoted as $\pi(\observations)$, is defined as the integral of the numerator in Eq. (\ref{eq Bayes'})
\begin{equation} \label{eq integral}
    \pi(\observations) = \int_{\mathbb{R}} L(\observations|\states) \pi_0(\states) d\states.
\end{equation}

The posterior density defined in Eq. (\ref{eq Bayes'}) represents the complete solution of the inverse problem in the Bayesian framework. The computation of the integral given in Eq. (\ref{eq integral}) presents challenges; nevertheless, it can be regarded as a normalizing constant. For the MCMC approach used in this paper, the value of this constant is irrelevant, as only relative values of the posterior probability are necessary. Thus, in the context of this paper, a posterior probability density function (PDF) lacking this normalizing factor is denoted as an \emph{unnormalized} PDF, represented as $\Tilde{\pi}(\states|\observations^*) = L(\observations^*|\states) \pi_0(\states)$ evaluated at the true measurement $\observations^*$.

In this particular problem, the prior is characterized as weakly informative, with the unknowns initially assumed to be uniformly distributed within a confined domain. This can be represented as:
\begin{equation} \label{eq prior}
    \pi_0 (\states) = \mathcal{U}_n (\states|B) ,
\end{equation}
where $B$ designates the bounded domain. Employing a uniform distribution as the prior specification guarantees an unbiased choice when there is little knowledge about the states. Furthermore, the likelihood PDF is defined as a multivariate normal distribution centered around the ground observation, $\observations^*$, with the covariance matrix characterized by independent and identically distributed (i.i.d) components, i.e. $\Sigma_\varepsilon = \sigma_\varepsilon^2 I$, where $\sigma_\varepsilon$ signifies measurement noise. The formulation can be expressed as follows
\begin{equation} \label{eq likelihood}
    L(\observations^*|\states) = \mathcal{N}(\observations^*|h(\states), \Sigma_\varepsilon) = \frac{1}{\sqrt{(2 \pi)^d \text{det} \Sigma_\varepsilon}} \exp \left( -\frac{1}{2} ||\observations^* - h(\states)||^2_{\Sigma_\varepsilon} \right) ,
\end{equation}
where the covariance-weighted norm is defined as
\begin{equation}
    ||\observations||^2_\Sigma = \observations^T \Sigma^{-1} \observations.
\end{equation}
Plugging Eqs. (\ref{eq prior}) and (\ref{eq likelihood}) into the expression for the unnormalized posterior distribution gives
\begin{equation}
    \Tilde{\pi}(\states|\observations^*) = \frac{\mathcal{U}_n}{\sqrt{(2 \pi)^d \text{det} \Sigma_\varepsilon}} \exp \left( -\frac{1}{2} ||\observations^* - h(\states)||^2_{\Sigma_\varepsilon} \right).
\end{equation}

In the context of Bayesian inference, we often find it more practical to work with the logarithmic representation of the equation provided earlier, as it offers advantages in terms of numerical stability related to machine-zero errors. The resulting logarithmic posterior disregarding the constants becomes
\begin{equation} \label{eq log-posterior}
    \Tilde{\pi}_l(\states|\observations^*) = - ||\observations^* - h(\states)||^2_{\Sigma_\varepsilon} + c_B(\states) ,
\end{equation}
where $c_B(\states)$ equals zero if state $\states$ lies within the domain $B$ and $-\infty$ otherwise. We will often define the domain $B$ to eliminate undesired states.
\subsubsection{Sampling and modeling of the posterior}
Samples are drawn from the log-posterior distribution given in Eq. (\ref{eq log-posterior}) using the Markov Chain Monte Carlo (MCMC) method \citep{metropolis1949monte,geman1984stochastic,hastings1970monte}. It generates a sequence of samples with each sample depending on the previous one, and the dependence is governed by a transition kernel that satisfies certain properties. In this paper, we adopt the Metropolis-Hastings (MH) algorithm as our choice of a transition kernel. Alternative algorithms like Gibbs sampling and Hamiltonian Monte Carlo exist and can be more efficient in exploring the state space when the Jacobian of the observation operator is available. However, MH remains a robust and simple choice, particularly where computing the Jacobian operator is not feasible or straightforward.
Since in the early stages of the MCMC algorithm, the chain starts from an arbitrary initial state and may not yet be representative of the target distribution, it is necessary to remove a portion of the initial chain entry known as burn-in to eliminate their influence on the target distribution.

The MH algorithm is very efficient in exploring a unimodal distribution. However, in complex multimodal distributions, the chain can get stuck in the local mode, making it difficult to explore the entire space. This decelerates convergence and can lead to difficulty in exploring the entire target distribution. To address this challenge, this study incorporates the concept of \emph{parallel tempering} (\cite{sambridge2014parallel}) (also known as \emph{replica exchange}) to improve the performance and convergence of the MCMC method. We have observed promising results by employing five Markov chains, each exploring the target distribution raised to respective powers of $5^p$, where $p$ ranges from -4 to 0 as integer values.

For the sake of simplicity in representing samples drawn from the MCMC algorithm, we approximate the generated samples by fitting them within another distribution denoted as $\pi_{\observations^*}(\states) \approx \pi(\states|\observations^*)$ which takes the form of a Gaussian Mixture Model (GMM) consisting of several Gaussian components. It is mathematically represented as a linear superposition of $K$ Gaussian distributions:
\begin{equation}
    \pi_{\observations^*}(\states) = \sum_{k=1}^K \alpha_k \mathcal{N} (\states|\bar{\states}^k, \Sigma^k).
\end{equation}
In this equation, $\alpha_k$ represents the weights that define the probability of a random variable $\states$ belonging to the Gaussian component $k$. It is worth noting that for the GMM to be a valid probability distribution, the sum of the weights must satisfy the condition:
\begin{equation}
    \sum_{k=1}^K \alpha_k = 1 .
\end{equation}

Throughout this paper, $\varepsilon=5 \times 10^{-4}$ is selected for the measurement noise unless otherwise stated. The algorithm begins with an initial $10^4$ step, using a diagonal variance of $10^{-4}$ for every state component in the proposal distribution. Subsequently, a $5 \times 10^{5}$ steps are executed with proposal variances uniformly set to $2.5 \times 10^{-5}$. The final sample dataset is extracted from the chain with $p = 0$, after discarding the first half of the chain as a burn-in period and retaining only every 100th chain entry from the remaining samples to reduce auto-correlations.

\section{Results and Discussion} \label{results}
In this section, we discuss the estimation results for various problems within an unbounded domain. We then explore the influence of boundary conditions to determine if they provide additional information and improve the estimation results. Temperature measurements are acquired from the observation model defined in Equation (\ref{eq forwardModel}), incorporating additive white noise with a zero-mean Gaussian distribution and variance $\sigma_\varepsilon^2$, $\varepsilon \sim \mathcal{N}(0,\sigma_\varepsilon^2)$. 

To generalize the problem, temperatures are represented relative to a reference temperature, denoted as $T_{ref}$. Unless stated otherwise, the sensors are aligned on $x$ axis within the range $[-1,1]$ with the measurement noise with variance equals to $5 \times 10^{-4}$. Due to symmetry, heat sources with state vectors $(x_o,y_o,q_h,c_1,c_2)$ and $(x_o,-y_o,q_h,c_1,c_2)$ yield identical temperature distributions at measurement locations. To avoid ambiguity, the estimator is constrained to positive $y$ regions via the bounding region in (\ref{eq log-posterior}). Additionally, in cases involving multiple heaters within the estimator, the entries in the state vector are rearranged in ascending order of $q_h$ at each MCMC step to eliminate relabeling symmetry. 

Various parameters influence the estimation process in heat conduction problems, such as the correlation between states, the number of measurements, sensor noise levels, the configurations of sensors and heaters, and the presence of boundaries. Each of these factors can significantly influence the estimation performance, emphasizing the necessity to investigate their impact on the estimation. We initially explore the steady-state heat conduction problem in an unbounded domain, deferring the consideration of boundary conditions to the latter part of this section.

The temperature field induced by a heater $\Omega_h$ in an unbounded medium is given by the Green's function solution of Eq. (\ref{eq heat}), 
\begin{equation}
    T(\posvec) = - q_h \iint_{\Omega_h} G(\posvec - \posvecheater) dA(\posvecheater) ,
    \label{eq Tconv}
\end{equation}
where $q_h$ has been assumed uniform and $dA=d \posvecheatcomp{1} d\posvecheatcomp{2}$ is the area of an infinitesimal element within the heater, and $G(\posvec)$ is the two-dimensional Green's function for the Laplacian,
\begin{equation}
    G(\posvec) = \frac{1}{2\pi} \log |\posvec|.
\end{equation}
This solution is easily generalized to a collection of heaters, of course.

\subsection{Multipole expansion of the temperature field}
For the purpose of gaining some insight into the sensitivity of measurements to states, we initiate our exploration by considering only a single heater and leveraging the multipole expansion of Green's function $G(\posvec - \posvecheater)$ about the center of the heater, taken for the purposes of this discussion as $\posvecheater=0$. (In other words, the observation location $\posvec$ should be regarded as the position relative to this center.) The multipole expansion is valid at the evaluation points far from the heater's center (compared to the size of the heater). We are making that assumption in this initial analysis only to illustrate the sensitivity of the temperature observations to heater parameters.

The multipole expansion is given by \citep{eldredge2019mathematical}
\begin{equation}
    \begin{aligned}
        2 \pi G(\posvec - \posvecheater) &= \log r - \posvecheatcomp{i} \frac{\partial}{\partial \posveccomp{i}} \log r + \frac{1}{2} \posvecheatcomp{i} \posvecheatcomp{j} \frac{\partial^2}{\partial \posveccomp{i} \partial \posveccomp{j}} \log r - \cdots \\
        &= \log r - \posvecheatcomp{i} \frac{\posveccomp{i}}{r^2} + \frac{1}{2} \posvecheatcomp{i} \posvecheatcomp{j} \left( \frac{\delta_{ij}}{r^2} - \frac{2 \posveccomp{i} \posveccomp{j}}{r^4} \right) - \cdots,
    \end{aligned}
\end{equation}
where summation over indices $i$ and $j$ are implied, and $r =|\posvec| = (r_x^2+r_y^2)^{1/2}$. Here $r_x$ and $r_y$ denote the distance between any point and the center of the heater in the $x$ and $y$ directions, respectively. Substituting this expansion into the convolution integral (\ref{eq Tconv}) leads to the expression:
\begin{align}
    T(\posvec) &= -\frac{q_h}{2 \pi} \left( \log r \iint_{\Omega_h} dA - \frac{\posveccomp{i}}{r^2} \iint_{\Omega_h}  \posvecheatcomp{i} dA + \right. \nonumber\\
    & \hspace{1.5cm} \left. \frac{1}{2} \left( \frac{\delta_{ij}}{r^2} - \frac{2 \posveccomp{i} \posveccomp{j}}{r^4} \right) \iint_{\Omega_h}  \posvecheatcomp{i} \posvecheatcomp{j} dA - \cdots \right).
\end{align}
The integral in the first term corresponds to the area of the heater, denoted as $A$. The integral in the second term represents the first moment of the heater and signifies the location of the heater's center, identically zero by definition. The integral in the third term represents the second moment tensor of the heater about its center, denoted by $M_{ij}$, and represents the size and basic shape of the heater. Applying these concepts, the temperature distribution can be simplified to
\begin{equation}
    T(\posvec) = -\frac{Q}{2 \pi} \log r - \frac{q_h}{4 \pi} M_{ij} \left( \frac{\delta_{ij}}{r^2} - \frac{2 \posveccomp{i} \posveccomp{j}}{r^4} \right) + O(1/r^3).
\end{equation}
As evident from this equation, the temperature field exhibits non-linearity concerning the relative position and linearity for the total heat generation $Q = q_hA$ and the size and shape of the heater, described by $M_{ij}$. However, the temperature influence from the size and shape decay rapidly with distance ($\sim r^{-2}$).

To facilitate further analysis, we can linearize the temperature field at every evaluation point, $\posvec_\alpha$; $\alpha=1,2,\cdots,d$, imposed by state x about that determined by the true state $\text{x}^*$, yielding 
\begin{equation}
    T(\posvec_\alpha,\states) \approx T(\posvec_\alpha,\states^*) + \pmb{H}_\alpha \cdot (\states - \states^*).
\end{equation}
Here, $\pmb{H}_\alpha \equiv \partial T(\posvec_\alpha,\states)/\partial \states |_{\states^*} \in \mathbb{R}^{1 \times n}$ represents the $\alpha^{th}$ row of the Jacobian of the observation operator at the true state. Each row of the Jacobian matrix signifies the sensitivity of a specific temperature sensor, $\alpha$, to the state vector. For a heater with a known shape, the partial derivatives in each row can be analytically computed as follows:
$[\partial T / \partial \posvecheatcomp{1} (= -\partial T / \partial r_x), \partial T / \partial \posvecheatcomp{2} (= -\partial T / \partial r_y), \partial T / \partial q_h]$. The Jacobian operator considering only the first two terms in the multipole expansion then becomes
\begin{equation} \label{eq Jacobian}
    \begin{aligned}
        \pmb{H}_\alpha = \left[ \vphantom{\frac{1}{2}} \right. & \left( \frac{Q}{2 \pi} \frac{\partial}{\partial r_{x}} \log \posveccomp{\alpha} + \frac{q_h}{4 \pi} M_{ij} \frac{\partial}{\partial r_{x}} \left( \frac{\delta_{ij}}{\posveccomp{\alpha}^2} - \frac{2 \posveccomp{\alpha,i} \posveccomp{\alpha,j}}{\posveccomp{\alpha}^4} \right) \right),\\
        & \left( \frac{Q}{2 \pi} \frac{\partial}{\partial r_{y}} \log \posveccomp{\alpha} + \frac{q_h}{4 \pi} M_{ij} \frac{\partial}{\partial r_{y}} \left( \frac{\delta_{ij}}{\posveccomp{\alpha}^2} - \frac{2 \posveccomp{\alpha,i} \posveccomp{\alpha,j}}{\posveccomp{\alpha}^4} \right) \right),\\
        & \left. \left( -\frac{A}{2 \pi} \log \posveccomp{\alpha} - \frac{1}{4 \pi} M_{ij} \left( \frac{\delta_{ij}}{\posveccomp{\alpha}^2} - \frac{2 \posveccomp{\alpha,i} \posveccomp{\alpha,j}}{\posveccomp{\alpha}^4} \right) \right) \right] ,
    \end{aligned}
\end{equation}
where $\alpha$ denotes the $\alpha^{th}$ sensor corresponding to the $\alpha^{th}$ row of $\pmb{H}$, and each column of $\pmb{H}$ signifies the sensitivity of all sensors to a component of the state vector.

By analyzing columns of $\pmb{H}$, one can predict several aspects of the estimation before performing the inference process. Intuitively, if a specific column $l$ in $\pmb{H}$ is significantly smaller than others, it implies that the sensors have minimal sensitivity to changes in the corresponding state component $\text{x}_l$. Consequently, $\text{x}_l$ will exhibit maximum uncertainty in the estimation. This insight aligns with an alternate approach involving the singular value decomposition of the Jacobian, $\pmb{H}=USV^T$. The greatest uncertainty was shown by \cite{eldredge2023bayesian} to be associated with the smallest singular value of $\pmb{H}$, $s_n$. The corresponding eigenvector $v_n$ indicates the mixture of states for which the estimation faces the most ambiguity.

The foregoing analysis will serve an important role in the following sections to justify some observed behavior of heaters. Note that while multipole expansion serves as a valuable tool for sensitivity analysis, its application for temperature estimation in the domain, especially with larger heaters, may yield less accurate results due to the growing importance of higher-order terms in the expansion. Now that we have some intuition about how a heat source behaves, we can begin our discussion on the results.

\subsection{Unbounded domain}
From hereon, we no longer make assumptions about the proximity of sensors to the heaters. The temperature field is influenced by the total heat generated within the domain as well as the location of the heater. This means that adjusting the heater's strength and shape simultaneously, while keeping the overall heat generation normalized by thermal conductivity given by $Q=q_hA$ constant, does not alter the temperatures at sensor locations. This conclusion is supported by the results obtained from assuming a circular heater with radius $c_1$ (setting $c_2=0$ in the Fourier series) and estimating state variables, $\states=(x_o,y_o,q_h,c_1)$, as follows. In this example, the true state is given as $\states=(0.5,0.8,1,0.5,0)$ and the goal is to estimate the states of the heater. Note that here $c_2$ is given as \emph{a priori} known, via a sharp Gaussian distribution around zero mean and variance of $10^{-6}$.

\begin{figure}[ht!]
\begin{subfigure}{0.5\textwidth}
    \centering
    \includegraphics[width=1\linewidth]{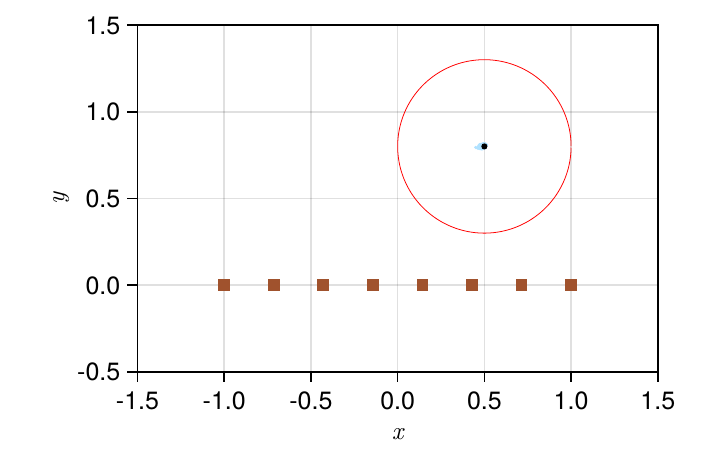} 
    \caption{}
    \label{fig:sample2sens}
\end{subfigure}
\begin{subfigure}{0.5\textwidth}
    \centering
    \includegraphics[width=1\linewidth]{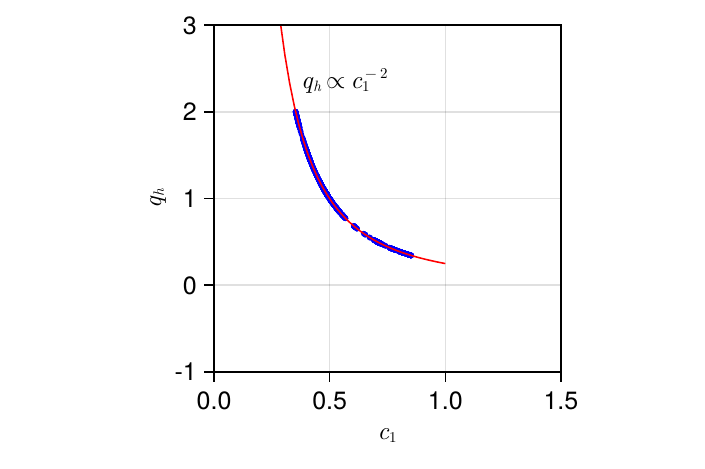} 
    \caption{}
    \label{fig:sample3sens}
\end{subfigure}

    \caption{\centering One true heater and one heater estimator with eight temperature sensors. The heater is given to be circular with radius $c_1$ and true state $(0.5,0.8,1,0.5,0)$; (a) samples generated with MCMC with the red curve representing the boundary of the heater, and (b) generated samples on $q_h-c_1$ plot.}
    \label{fig:circle}
\end{figure}

\begin{figure}[ht!]
\begin{subfigure}{0.3\textwidth}
    \centering
    \includegraphics[width=1\linewidth]{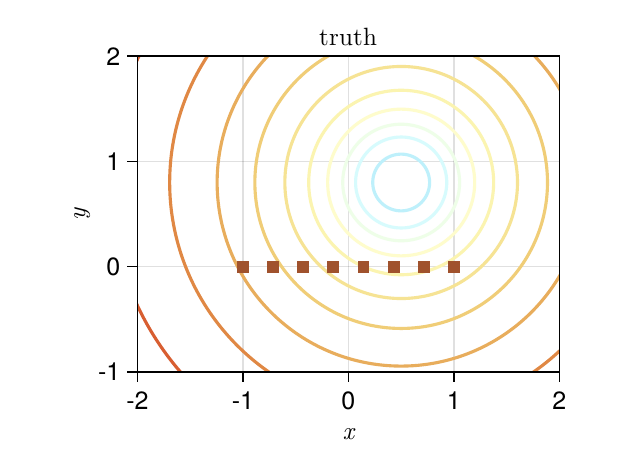} 
    \caption{}
    \label{fig:}
\end{subfigure}
\begin{subfigure}{0.3\textwidth}
    \centering
    \includegraphics[width=1\linewidth]{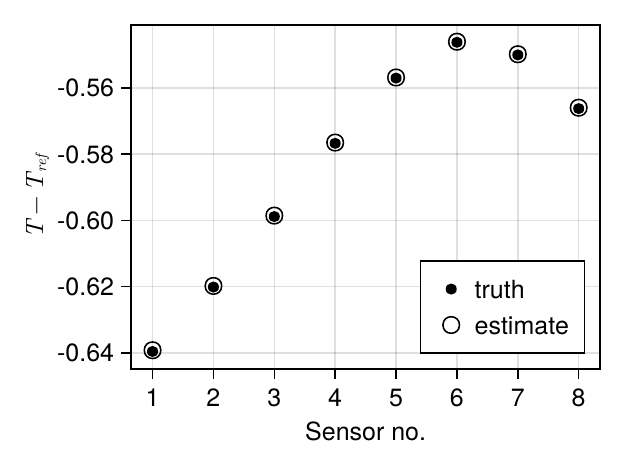} 
    \caption{}
    \label{fig:}
\end{subfigure}
\begin{subfigure}{0.3\textwidth}
    \centering
    \includegraphics[width=1\linewidth]{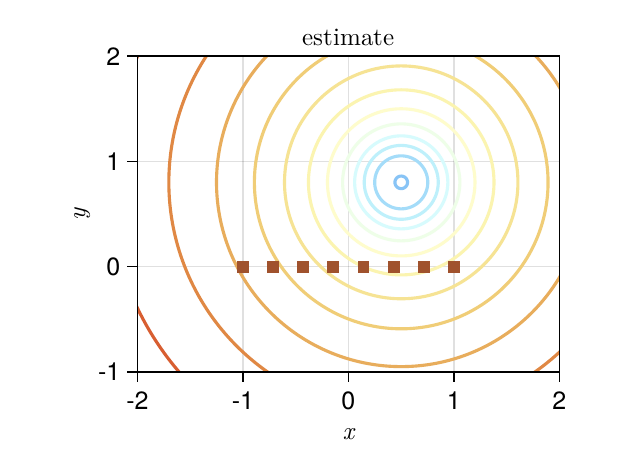} 
    \caption{}
    \label{fig:}
\end{subfigure}

    \caption{\centering One true heater and one heater estimator with eight temperature measurements characterized by five Gaussian components. The heater is given to be a circle with radius $c_1$ and true states $(0.5,0.8,1,0.5,0)$. (a) true temperature contours, (b) temperature distribution at the sensor locations, and (c) estimated temperature contours.}
    \label{fig:TempCircle}
\end{figure}

Figure \ref{fig:circle}(a) illustrates the estimated position of the center of the heater with close proximity to the true state represented with a black dot. The true shape of the heater is also depicted in this figure with a red line. In contrast to the success in estimating the heater's position, Figure \ref{fig:circle}(b) reveals a strong correlation between the heater's strength and its radius, evidenced by the curve fit $q_h c_1^2=C$ where $C$ is a constant. This correlation is a fundamental characteristic; in the context of a circular heater with radius $c_1$ and uniform strength $q_h$, the total heat generated in the domain is denoted as $Q=q_hA=\pi q_h c_1^2$. Consequently, maintaining the total heat generation constant, i.e. a constant value for $q_h c_1^2$, while adjusting $q_h$ and $c_1$ results in consistent temperature patterns around the heater and an equally probable combination of states for strength and shape. This behavior is further illustrated in Figure \ref{fig:TempCircle}. The estimation in this figure corresponds to the mean of the Gaussian mixture component with the highest likelihood, which is $\states = (0.49,0.8,1.91,0.36,0)$.

Hence, to ensure accurate inference in proximity to the true states, it is advisable to refrain from simultaneously estimating both $q_h$ and the heater's size due to their indistinguishability. For the current analysis, we first assume the heater's shape and size to be known with a high degree of confidence, focusing solely on estimating the position of its center as well as its strength (three states to be estimated). This prior knowledge about any known state is incorporated into the estimation framework by imposing sharp Gaussian priors centered around their true values with variance given as $10^{-6}$. We then assess the position and shape inference, giving the heater's strength as known. For the single heater estimation, the heater is centered at $(x_o,y_o) = (0.5,0.8)$ with unit strength and shape coefficients $(c_1,c_2) = (0.5,0.25)$, unless otherwise stated.
\subsubsection{Effect of the number and arrangement of sensors}
\begin{figure}[ht!]
\begin{subfigure}{0.5\textwidth}
    \centering
    \includegraphics[width=1\linewidth]{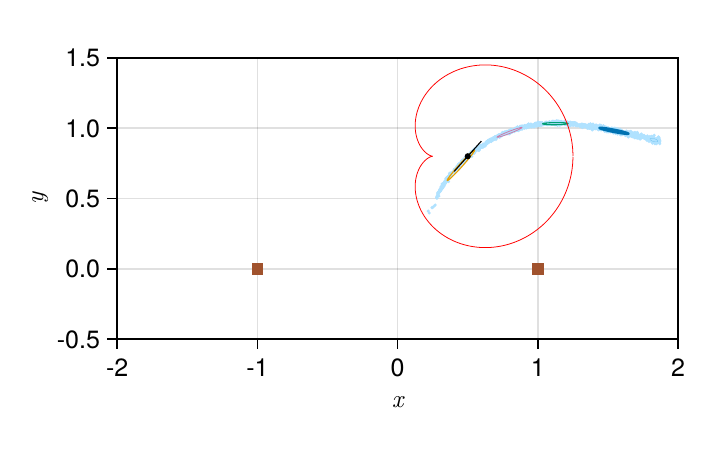} 
    \caption{}
    \label{fig:sample2sens}
\end{subfigure}
\begin{subfigure}{0.5\textwidth}
    \centering
    \includegraphics[width=1\linewidth]{samples-1heater-1truth-K5-3sensor-ds100-noise5em4.pdf} 
    \caption{}
    \label{fig:sample3sens}
\end{subfigure}
\begin{subfigure}{1\textwidth}
    \centering
    \includegraphics[width=0.5\linewidth]{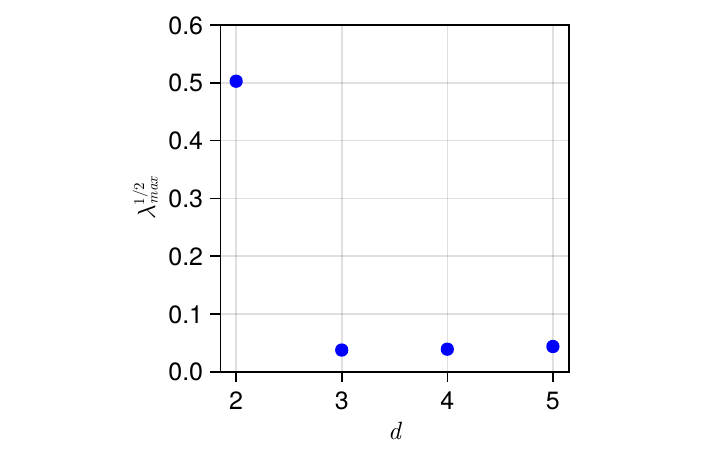} 
    \caption{}
    \label{fig:eig_sens}
\end{subfigure}

    \caption{\centering One true heater and one heater estimator characterized by five Gaussian components with (a) two and (b) three temperature measurements. The blue dots on the graph represent samples drawn from MCMC. The black circle indicates the true position of the heater’s center, while the black line illustrates the direction of maximum uncertainty at the true heater location. The boundary of the heater is shown with a red curve. (c) the maximum uncertainty for varying number of sensors.}
    \label{fig:numnerMeasurements}
\end{figure}

The number of sensors is expected to have a significant effect on the accuracy of the inference problem. Well-distributed sensors covering different regions of the domain can provide a more comprehensive understanding of the temperature field. If the number of sensors is less than the states to be estimated, the problem becomes rank deficient, indicating that the available information is insufficient to uniquely identify all the states. This conclusion becomes evident in Figure \ref{fig:numnerMeasurements}, where it is shown that two measurements are inadequate for accurate state estimation, whereas three measurements prove sufficient. In the case of two temperature sensors, the MCMC samples are distributed along an arc, holding equally probable states in the Euclidean space.

The analysis of the covariance matrix for systems with different numbers of measurements provides valuable insights into estimation accuracy. Principal Component Analysis (PCA), a prominent statistical technique, is employed for dimensionality reduction while preserving essential patterns. PCA utilizes eigenvalue decomposition on the covariance matrix, where eigenvectors signify directions of maximum uncertainty in the data, and the square root of eigenvalues denotes the magnitudes of deviation along these directions. Specifically, the largest eigenvalue of the covariance matrix corresponds to the greatest variance in the direction with the highest uncertainty. 

The covariance matrix of the Gaussian mixture component with the highest likelihood is selected for analysis. Comparing the square root of the maximum eigenvalues of the covariance matrices with two and three sensors in Figure \ref{fig:numnerMeasurements}(c) reveals a substantial drop, indicating a significant improvement in estimation precision with the addition of the third sensor. Interestingly, beyond three sensors, the maximum eigenvalue shows minimal variation, signifying that adding more sensors does not lead to a substantial enhancement in the system's overall estimation.

\begin{figure}[ht!]
\begin{subfigure}{0.3\textwidth}
    \centering
    \includegraphics[width=1\linewidth]{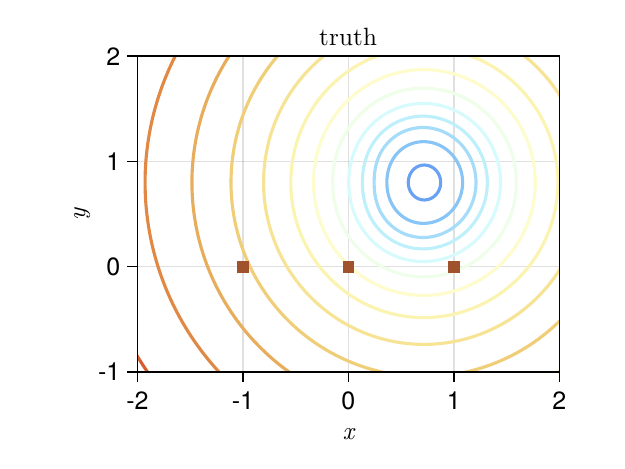} 
    \caption{}
    \label{fig:}
\end{subfigure}
\begin{subfigure}{0.3\textwidth}
    \centering
    \includegraphics[width=1\linewidth]{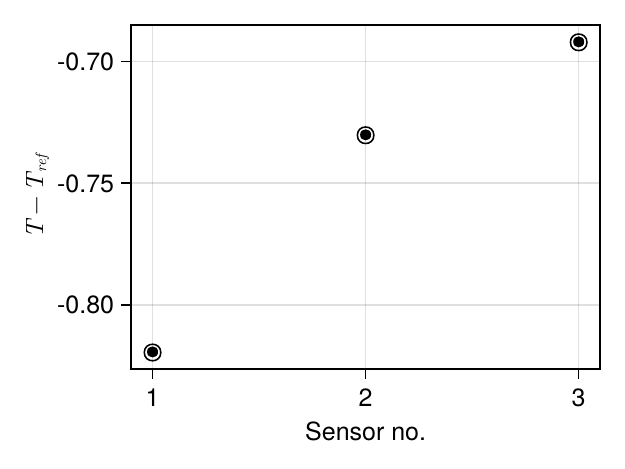} 
    \caption{}
    \label{fig:}
\end{subfigure}
\begin{subfigure}{0.3\textwidth}
    \centering
    \includegraphics[width=1\linewidth]{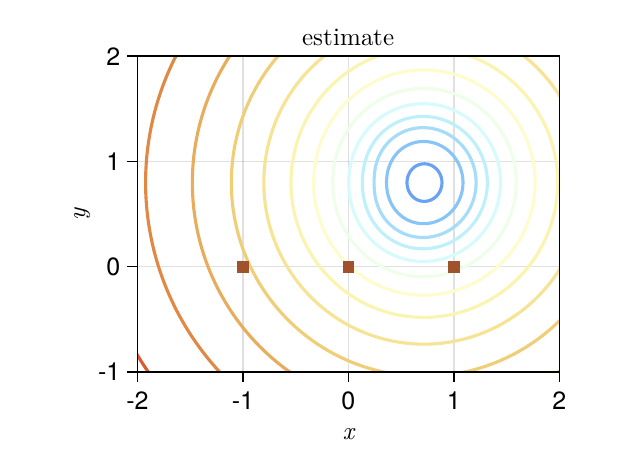} 
    \caption{}
    \label{fig:}
\end{subfigure}

    \caption{\centering One true heater and one heater estimator with three temperature measurements characterized by five Gaussian components. The true state is given as $(0.5,0.8,1,0.5,0.25)$ with \emph{a priori} known shape. (a) true temperature contours, (b) temperature distribution at the sensor locations, and (c) estimated temperature contours.}
    \label{fig:Temp3sens}
\end{figure}

Figure \ref{fig:Temp3sens}(a),(c) display the true as well as estimated temperature contours for the case of three sensors positioned along the $x$ axis. The calculated estimated states, $(0.5,0.8,1.01,0.5,0.25)$, closely align with the actual states, indicating a high level of accuracy in the estimation process. Additionally, Figure \ref{fig:Temp3sens}(b) provides a comparison of temperature distributions at the measurement locations, clearly illustrating the remarkable agreement between the temperatures derived from both the truth and the estimation results. This consistency underscores the effectiveness of the estimation approach in capturing the intricate temperature patterns within the system.

\begin{figure}[ht!]
    \centering
    \includegraphics[width=0.5\textwidth]{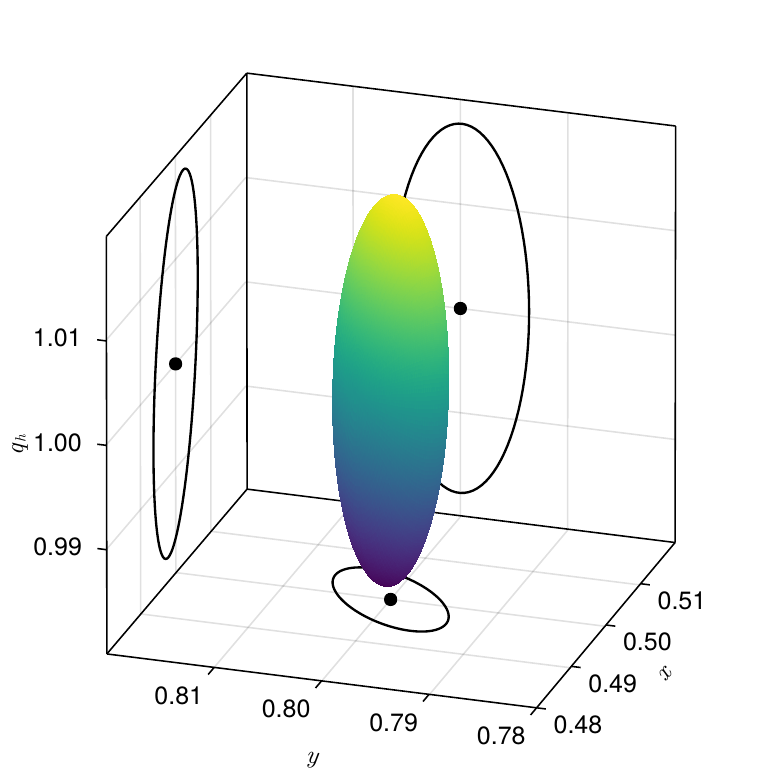}
    \caption{\centering Direction of maximum uncertainty for one true heater and one heater estimator with three temperature measurements.}
    \label{fig:ellipsoid}
\end{figure}

Figure \ref{fig:ellipsoid} shows the ellipsoid for the covariance matrix belonging to the Gaussian component with the highest likelihood centered at the true state. This figure indicates that most of the uncertainty is in the direction of $q_h$, proving the fact that a wider range of heat strength will give temperature results in the specified sensor uncertainty at the measurement locations. The direction of greatest uncertainty is calculated to be $(x_o,y_o,q_h)=(-0.13,-0.003,-0.95)$.

Consider that due to the fundamental principles of the temperature Poisson equation, heat diffuses away from the source in a semi-circular pattern. Near the heat source, the temperature contours might be more concentrated and distorted, reflecting the intricate geometry of the heat source. As the distance from the source increases, the contours tend to smooth out and become more uniform. This phenomenon is pictured in Figure \ref{fig:config}(b) showcasing three non-circular heaters. Away from these heat sources, the temperature contours form a smooth pattern, obscuring the shapes of the heaters. As such, if sensors are arranged in a circular formation centered at the heater, the temperature difference between sensors becomes minimal, providing limited information about the system. Instead, selecting sensor arrangements that capture a wider range of temperature variations is crucial for enhancing the quality of the estimation. In pursuit of this, sensors are aligned along the $x$ axis in the rest of this paper.

\subsubsection{Effect of heater size}
\begin{figure}[htp!]
\begin{subfigure}{0.5\textwidth}
    \centering
    \includegraphics[width=1\linewidth]{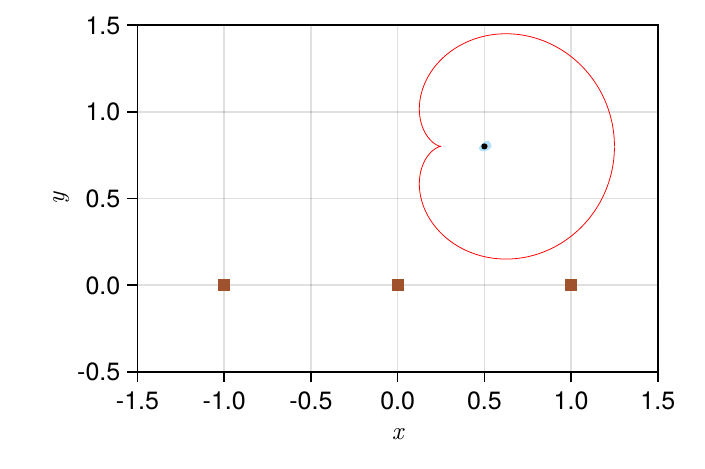} 
    \caption{}
    \label{fig:largeHeater}
\end{subfigure}
\begin{subfigure}{0.5\textwidth}
    \centering
    \includegraphics[width=1\linewidth]{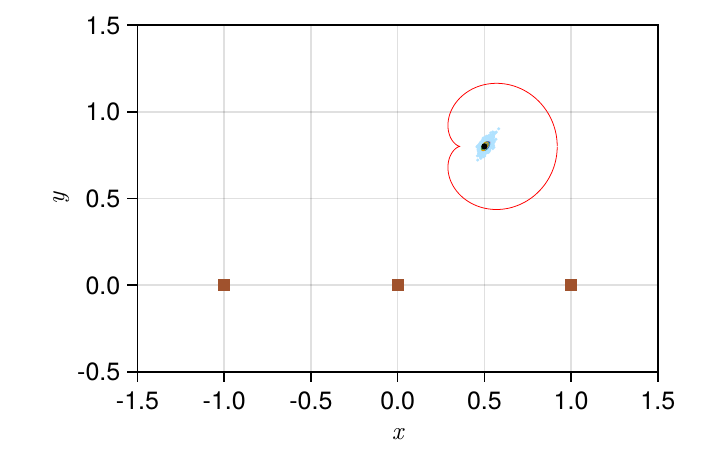} 
    \caption{}
    \label{fig:mediumHeater}
\end{subfigure}
\begin{subfigure}{0.5\textwidth}
    \centering
    \includegraphics[width=1\linewidth]{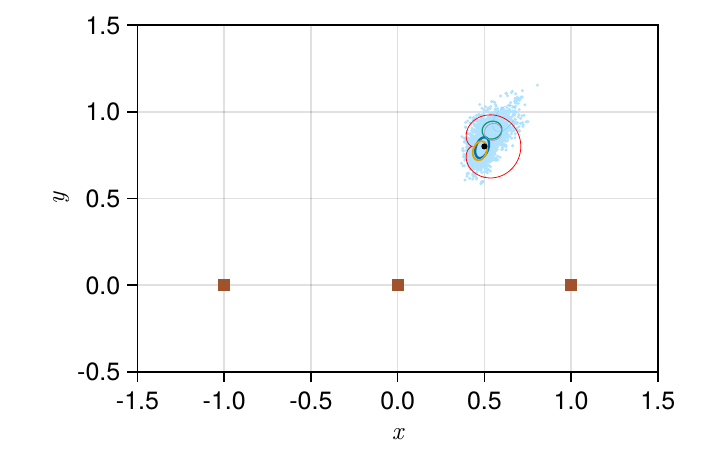} 
    \caption{}
    \label{fig:smallHeater}
\end{subfigure}
\begin{subfigure}{0.5\textwidth}
    \centering
    \includegraphics[width=1\linewidth]{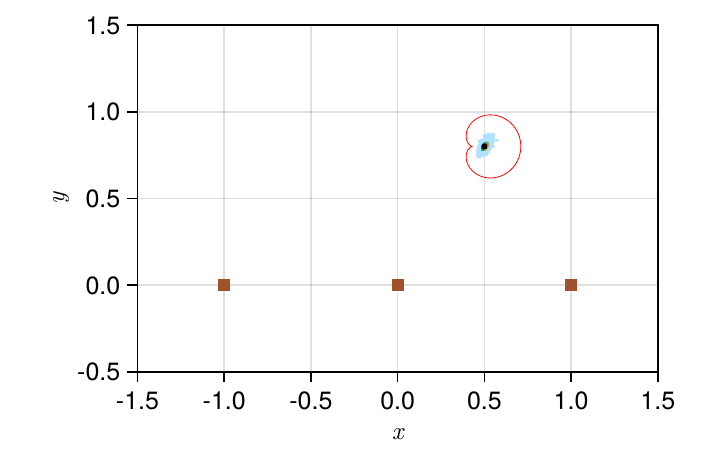} 
    \caption{}
    \label{fig:eig_heater_size}
\end{subfigure}
\begin{subfigure}{0.5\textwidth}
    \centering
    \includegraphics[width=1\linewidth]{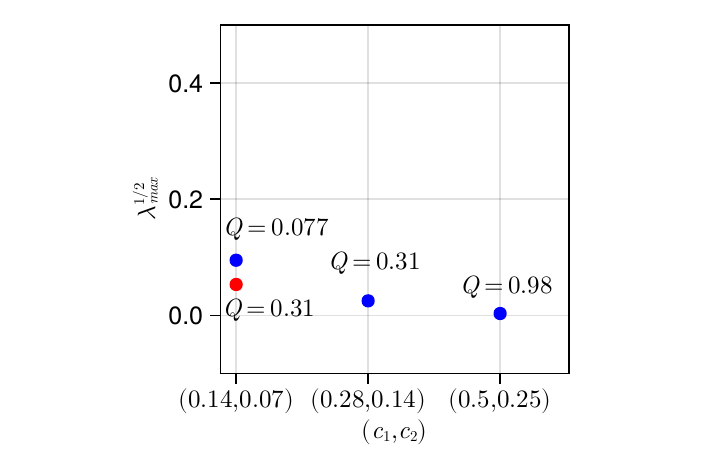} 
    \caption{}
    \label{fig:eig_heater_size}
\end{subfigure}
\begin{subfigure}{0.5\textwidth}
    \centering
    \includegraphics[width=1\linewidth]{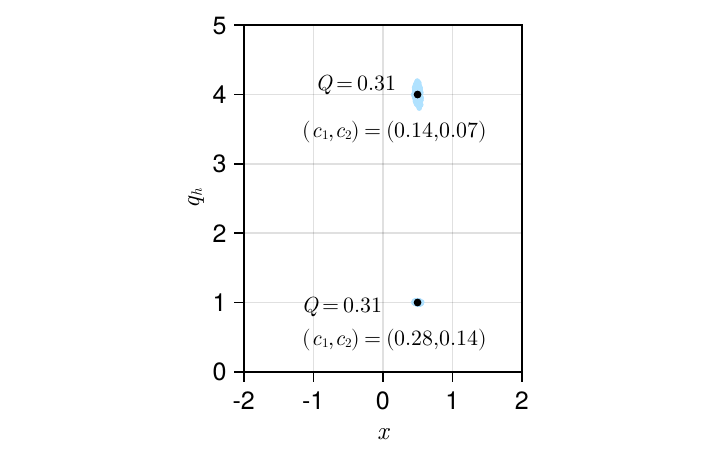} 
    \caption{}
    \label{fig:eig_heater_size}
\end{subfigure}
    \caption{\centering One true heater and one heater estimator with three temperature measurements, characterized by five Gaussian components for different heater sizes (a) $(c_1,c_2)=(0.5,0.25)$, (b) $(c_1,c_2)=(0.28,0.14)$, (c) $(c_1,c_2)=(0.14,0.07)$. The true strength for cases (a)-(c) is equal to $q_h=1$ corresponding to the total heat generation of $Q=0.98$, $0.31$, and $0.077$ respectively. (d) estimation results for $(c_1,c_2)=(0.14,0.07)$ with $Q=0.31$. The red curve in all cases illustrates the boundary of the heater. (e) maximum uncertainty for heaters corresponding to cases (a)-(d). The blue circles correspond to cases (a)-(c) where the true strength is the same, while the red circle is associated with the case (d) where the total heat is kept consistent. (f) samples on $x-q_h$ plot for cases (b) and (d). Black circles in all cases represent the true state.}
    \label{fig:heaterSize}
\end{figure}

Understanding how the physical dimensions of the heater influence the reliability and accuracy of the inferred states is essential for robust and trustworthy results. In this context, we focus specifically on the variation in heater size while maintaining the center and heat strength constant at $(x_o,y_o,q_h)=(0.5,0.8,1)$. Figure \ref{fig:heaterSize}(a)-(c) depicts the samples generated using MCMC for heaters with varying sizes but consistent shapes, by maintaining a constant ratio of $c_1/c_2=2$. The outcomes together with the maximum length of uncertainty depicted in Figure \ref{fig:heaterSize}(e) (blue circles) indicate greater uncertainty in estimation for the smaller heater. As anticipated, inference accuracy improves when the boundary of the heater is closer to the sensors, underscoring the critical role of sensor proximity in enhancing the quality of estimation results. Furthermore, maintaining a consistent strength while decreasing the heater's size translates to a reduced total heat generated in the domain (refer to the total heat generation indicated near each circle in Figure \ref{fig:heaterSize}(e)). Consequently, temperatures at the measurement locations decrease correspondingly. This reduction in temperature amplifies the ratio of sensor noise to the temperature at its location, leading to a degradation in estimation performance as heater shrinks.

In another scenario, the size of the heater was reduced from $(c_1,c_2)=(0.28,0.14)$ to $(c_1,c_2)=(0.14,0.07)$ while maintaining its total heat generation constant at $Q=0.31$. The results depicted in Figure \ref{fig:heaterSize}(b),(d) clearly demonstrate that reducing the heater's dimensions while keeping its total heat constant has minimal effect on estimating the center of the heaters. However, the heater's size significantly influences the uncertainty in $q_h$. Figure \ref{fig:heaterSize}(e) provides visual evidence supporting this observation. The amplified uncertainty associated with the smaller heater manifests in the augmented sample variances portrayed in Figure \ref{fig:heaterSize}(f) for $q_h$. Eq. (\ref{eq Jacobian}) provides an explanation for this observation. In this equation, the last column of the Jacobian matrix corresponds to the sensitivity of measurements to strength $q_h$. When considering two heaters with consistent $Q$ but differing sizes, it becomes evident that the last column of the Jacobian matrix $\pmb{H}$ is smaller for the smaller heater. This is attributed to the reduced area $A$ and moment $M_{ij}$ associated with the smaller heater in comparison to its larger counterpart. This observation highlights that the smaller heater exhibits lower sensitivity to $q_h$ than the larger one. Consequently, this disparity results in increased uncertainty in estimating $q_h$ for the smaller heater, as visually depicted in Figure \ref{fig:heaterSize}(f).

\subsubsection{Effect of sensor noise}
\begin{figure}[ht!]
\begin{subfigure}{0.5\textwidth}
    \centering
    \includegraphics[width=1\linewidth]{figures/samples-1heater-1truth-K5-3sensor-ds100-noise5em4.pdf} 
    \caption{}
    \label{fig:noise5em4}
\end{subfigure}
\begin{subfigure}{0.5\textwidth}
    \centering
    \includegraphics[width=1\linewidth]{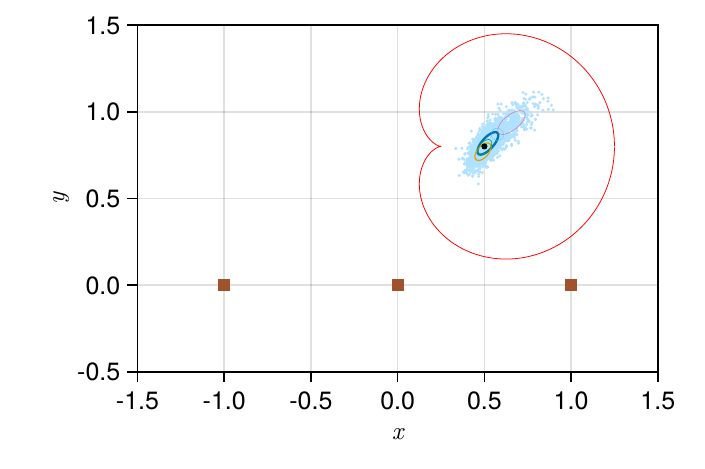} 
    \caption{}
    \label{fig:noise5em3}
    
\end{subfigure}
\begin{subfigure}{1\textwidth}
    \centering
    \includegraphics[width=0.5\linewidth]{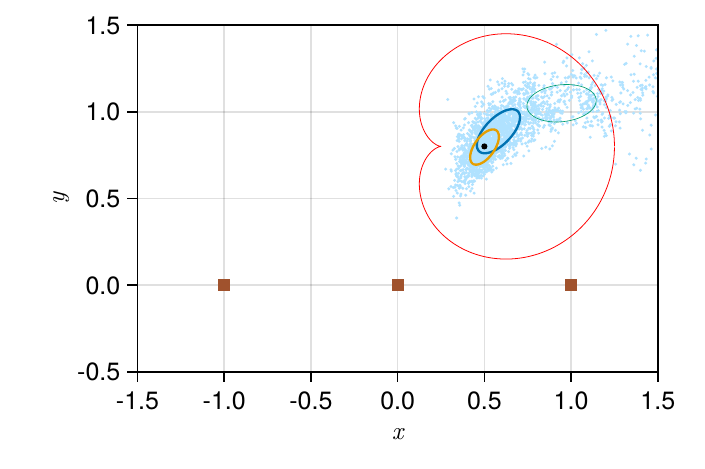} 
    \caption{}
    \label{fig:noise1em2}
\end{subfigure}

    \caption{\centering One true heater and one heater estimator with three temperature measurements, characterized by five Gaussian components for sensor noise equals to (a) $5\times 10^{-4}$, (b) $5\times 10^{-3}$, and (c) $1\times 10^{-2}$. The red curve illustrates the boundary of the heater.}
    \label{fig:noise}
\end{figure}

In this section, we investigate the impact of sensor noise levels on the accuracy of estimation. One true heater and one heater estimator are used for this part while the true states of the heater are set at $(x_o,y_o,q_h,c_1,c_2)=(0.5,0.8,1,0.5.0.25)$. The noise level denoted as $\sigma_\varepsilon$ varies from $5 \times 10^{-4}$ to $5 \times 10^{-3}$ and further to $1 \times 10^{-2}$. The findings in Figure \ref{fig:noise} reveal that as the noise level increases, samples become more dispersed, particularly away from sensors. These results indicate a systematic deviation from the true values, suggesting a biased error in estimation as the noise level escalates. The analysis of these biased errors provides valuable insights into the limitations of the estimation method under higher noise conditions.

\subsubsection{Estimating the shape}
\begin{figure}[ht!]
\begin{subfigure}{0.3\textwidth}
    \centering
    \includegraphics[width=1\linewidth]{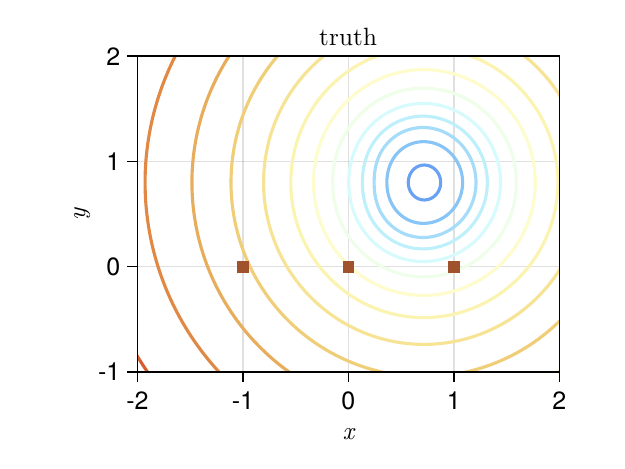} 
    \caption{}
    \label{fig:}
\end{subfigure}
\begin{subfigure}{0.3\textwidth}
    \centering
    \includegraphics[width=1\linewidth]{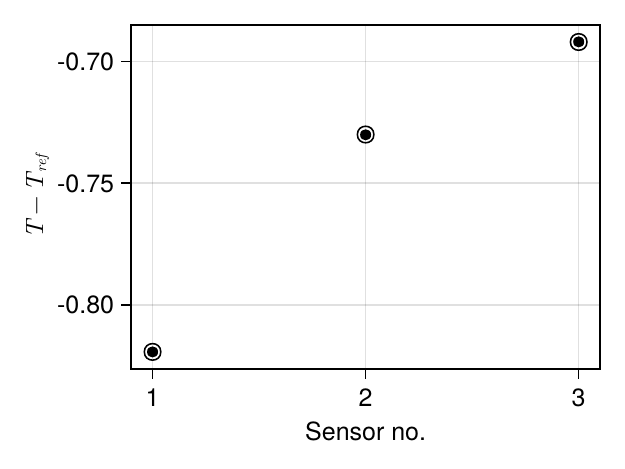}  
    \caption{}
    \label{fig:}
    
\end{subfigure}
\begin{subfigure}{0.3\textwidth}
    \centering
    \includegraphics[width=1\linewidth]{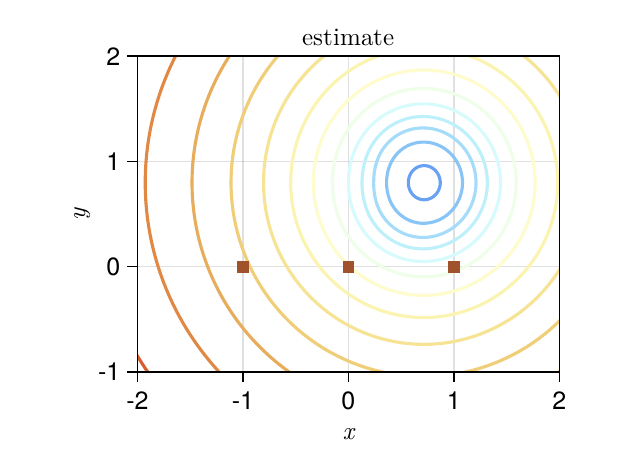}
    \caption{}
    \label{fig:}
\end{subfigure}

    \caption{\centering One true heater and one heater estimator with three temperature measurements. The position and strength of the heater are known and its shape is inferred. The true state is $(0.5,0.8,1,0.5,0.25)$.}
    \label{fig:known_r_q}
\end{figure}

\begin{figure}[ht!]
\begin{subfigure}{0.5\textwidth}
    \centering
    \includegraphics[width=1\linewidth]{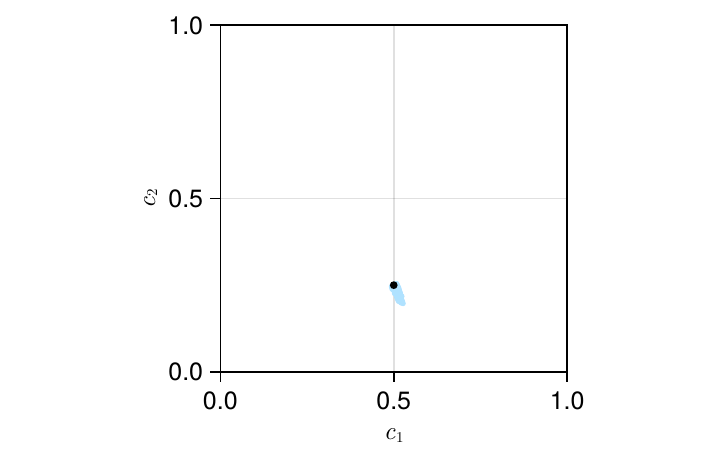} 
    \caption{}
    \label{fig:}
\end{subfigure}
\begin{subfigure}{0.5\textwidth}
    \centering
    \includegraphics[width=1\linewidth]{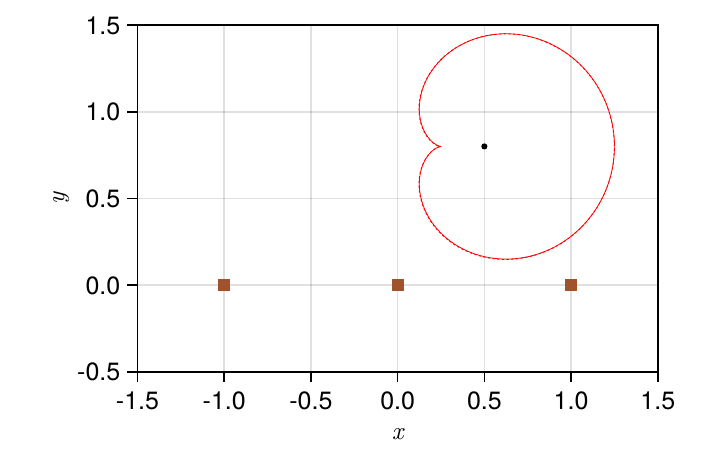} 
    \caption{}
    \label{fig:}
    
\end{subfigure}

    \caption{\centering One true heater and one heater estimator with three temperature measurements. The position and strength of the heater are known and its shape is inferred. (a) samples on $c_1-c_2$ plot and (b) the red curve illustrates the boundary of the heater, with the solid line representing the true shape and the dashed line indicating the estimated shape. The black circles illustrate the true state.}
    \label{fig:known_r_q_shape}
\end{figure}

In the context of shape estimation, the positions and strengths of the heaters are given as \emph{a priori} known, and the focus is on estimating their shapes. Figures \ref{fig:known_r_q} and \ref{fig:known_r_q_shape} showcase remarkably accurate results, aligning closely with the true shapes. Notably, the boundaries of the true and estimated heaters precisely overlap, as depicted in Figure \ref{fig:known_r_q_shape}(b).

\subsubsection{Estimating position and shape}
\begin{figure}[ht!]
\begin{subfigure}{0.5\textwidth}
    \centering
    \includegraphics[width=1\linewidth]{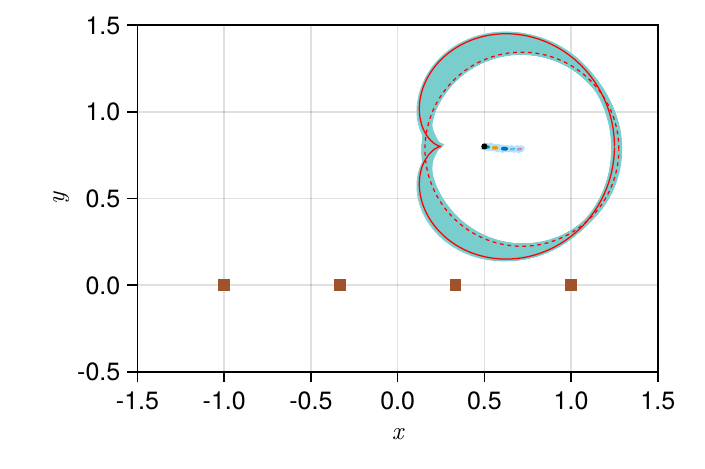} 
    \caption{}
    \label{fig:}
\end{subfigure}
\begin{subfigure}{0.5\textwidth}
    \centering
    \includegraphics[width=1\linewidth]{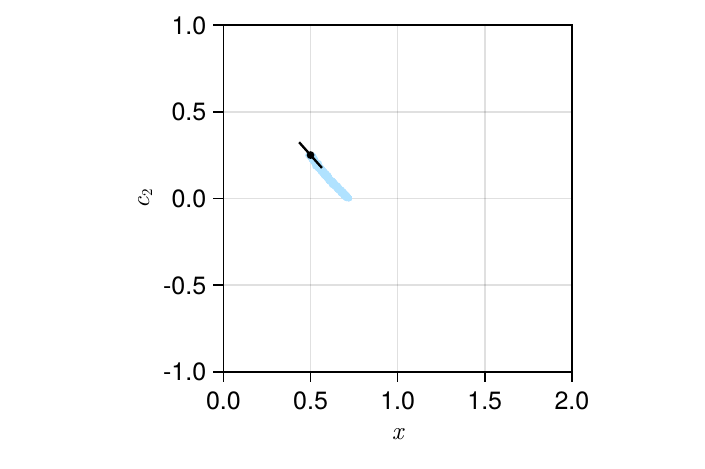} 
    \caption{}
    \label{fig:}
    
\end{subfigure}

    \caption{\centering One true heater and one heater estimator with four temperature measurements. The heater's strength is known, while its position and shape are inferred. (a) Samples generated with MCMC, represented by five Gaussian components. The boundaries of the heater shape for the true state and for the Gaussian component corresponding to the state with the most deviation from the true state are depicted by solid and dashed red curves, respectively. The shaded region shows the uncertainty envelop for shape estimation. (b) samples on $x-c_2$. The black dots represent the true state and the black line illustrates the direction and length of maximum uncertainty which mixes $x$ and $c_2$.}
    \label{fig:known_q}
\end{figure}

In certain applications, the generated heat strength within a domain is known, and the objective is to determine both the position and shape of the heating region. In this scenario, the unknown state vector consists of the parameters $(x_o, y_o, c_1, c_2)$. The true state is given as $(x_o,y_o,q_h,c_1,c_2)=(0.5,0.8,1.0,0.5,0.25)$ in this section. The outcomes of this estimation process are illustrated in Figure \ref{fig:known_q}. Figure \ref{fig:known_q}(a) displays the MCMC-generated samples alongside the true heater shape as a solid red curve, with the black circle denoting the true state. This figure reveals a precise estimation of the heater's position along the $y$-axis. Notably, excellent accuracy is also observed for the variable $c_1$. However, the samples exhibit dispersion from the true position, primarily in the positive $x$ direction. By employing PCA, the direction of maximum uncertainty is determined to be a combination of $x$ and $c_2$. Figure \ref{fig:known_q}(b) demonstrates the MCMC samples plotted on the $x-c_2$ plane, highlighting the direction of maximum uncertainty in the estimation process.

This observed behavior can be elucidated by examining the estimated shape corresponding to the Gaussian component displaying the most deviation from the true state. As illustrated by the red dashed curve in Figure \ref{fig:known_q}(a), this shape appears semi-circular, with a radius very close to the true $c_1$. However, its center is positioned away from the true location in the $x$ direction. The uncertainty envelope of the shape corresponding to all generated MCMC samples is pictured in Figure \ref{fig:known_q}(a) as a shaded green region. Notably, in the Fourier representation of the shape defined in Eq. (\ref{eq shape}), the term $c_2 e^{i2\theta}$ introduces a second harmonic component to the shape's boundary. When $c_2$ is non-zero, this term creates a deformation in the shape, deviating from a circle with radius $c_1$. As elucidated in previous sections, temperature contours do not preserve the deformed shape of the source away from the source itself. Consequently, the contours become circular away from the source, explaining the consistent temperature distribution observed at the sensor locations for the two heater shapes depicted in Figure \ref{fig:known_q}(a). Therefore, when inferring the heater's position and shape, the achieved accuracy is limited.

\subsubsection{Inference of multiple heating regions}
The preceding sections have primarily addressed the estimation of a single heater, assuming certain combinations of variables among the center's location, strength, and shape as \emph{a priori} known. In this section, we shift our focus to the inference of multiple heaters, where the shapes are assumed to be \emph{a priori} known and thus assigned sharp Gaussian priors. As a specific example, we consider the scenario where two heaters are placed within the domain. The inference problem involves determining the positions of the heaters' centers and their respective strengths. It is important to note that in the case of multiple heaters, the inverse problem may yield more than one plausible solution and it is worth statistically looking into them. The true heaters consist of $(x_{0,1},y_{o,1},q_{h,1},c_{1,1},c_{2,1})=(0.5,0.8,1.0,0.28,0.14)$ and $(x_{0,2},y_{o,2},q_{h,2},c_{1,2},c_{2,2})=(-0.6,0.6,2.0,0.2,0.0)$ with measurement noise equals to $\sigma_\varepsilon = 5 \times 10^{-4}$ as for the previous section. Various number of sensor measurements are used to investigate their effect on the inference performance.

\begin{figure}[ht!]
\begin{subfigure}{0.5\textwidth}
    \centering
    \includegraphics[width=1\linewidth]{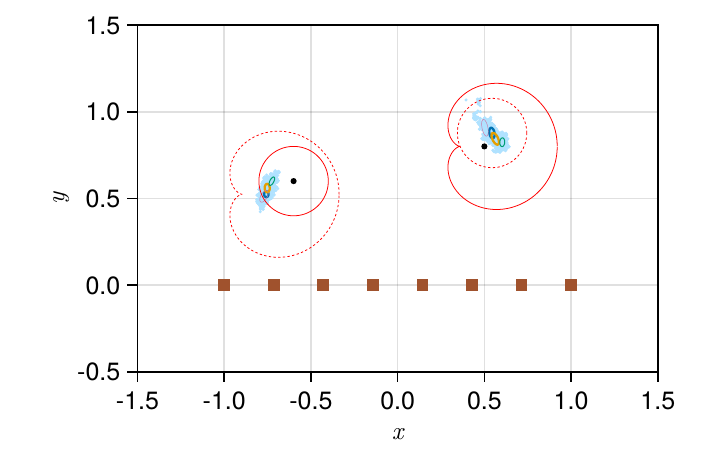} 
    \caption{}
    \label{fig:}
\end{subfigure}
\begin{subfigure}{0.5\textwidth}
    \centering
    \includegraphics[width=1\linewidth]{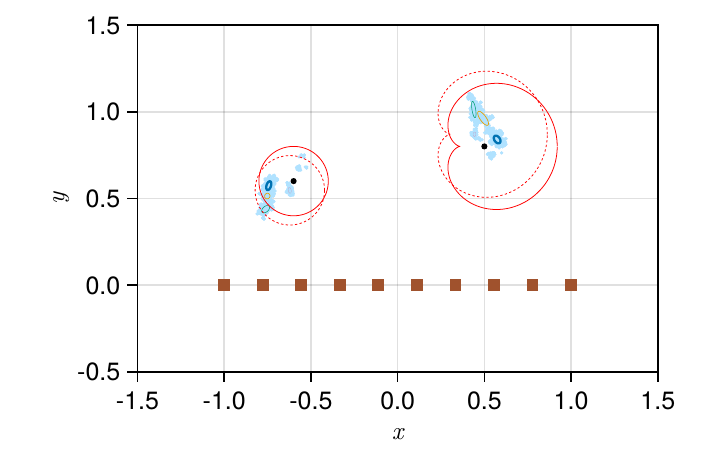} 
    \caption{}
    \label{fig:}
\end{subfigure}
\begin{subfigure}{1\textwidth}
    \centering
    \includegraphics[width=0.5\linewidth]{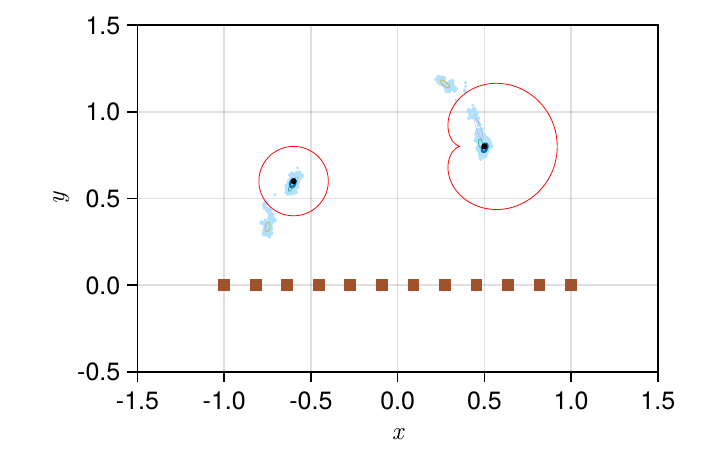} 
    \caption{}
    \label{fig:}
    
\end{subfigure}

    \caption{\centering Two true heater and two heater estimators with varying numbers of measurements: (a) 8 sensors, (b) 10 sensors, and (c) 12 sensors. The heaters' shapes are known, while their positions and strengths are inferred. The solid red curves represent the ground truth heaters, while the dashed shapes depict the estimated heaters.}
    \label{fig:twoHeaters_sensors}
\end{figure}

The MCMC results are presented in Figure \ref{fig:twoHeaters_sensors} for scenarios involving 8, 10, and 12 temperature measurements. The findings reveal that an increase in the number of temperature measurements, from 8 to 12, leads to a higher convergence rate toward the true state. In contrast to the cases with 8 and 10 sensors, the configuration with 12 sensors identifies the true states as the most probable states, exhibiting a greater likelihood.
Upon comparing the true and estimated heaters in Figure \ref{fig:twoHeaters_sensors}(a), it becomes evident that the inference with 8 sensors results in an inversion of the two heaters. This inversion is attributed to assigning a very high value of $q_h$ to the estimated circular heater and a very low $q_h$ to the larger heart-shaped heater, leading to a similar temperature distribution at the sensor locations. As the number of measurements increases from 8 to 10 in \ref{fig:twoHeaters_sensors}(b), the correct order of heaters is established, and the exploration of the state space improves, generating several samples close to the true states. However, only the scenario with 12 sensors manages to produce a sufficient number of samples precisely at the true states associated with the greatest likelihood. 

\begin{figure}[ht!]
\begin{subfigure}{0.3\textwidth}
    \centering
    \includegraphics[width=1\linewidth]{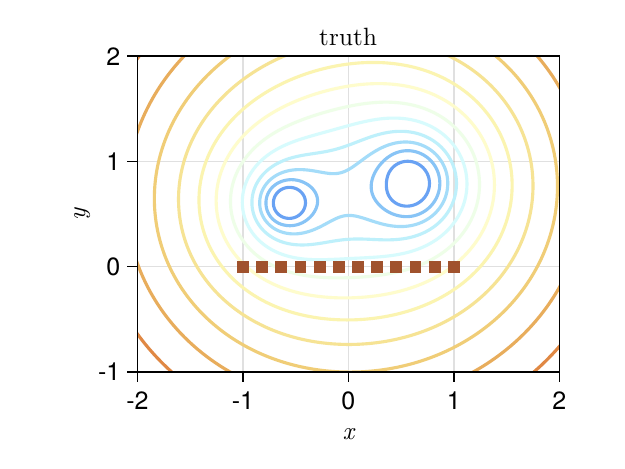} 
    \caption{}
    \label{fig:}
\end{subfigure}
\begin{subfigure}{0.3\textwidth}
    \centering
    \includegraphics[width=1\linewidth]{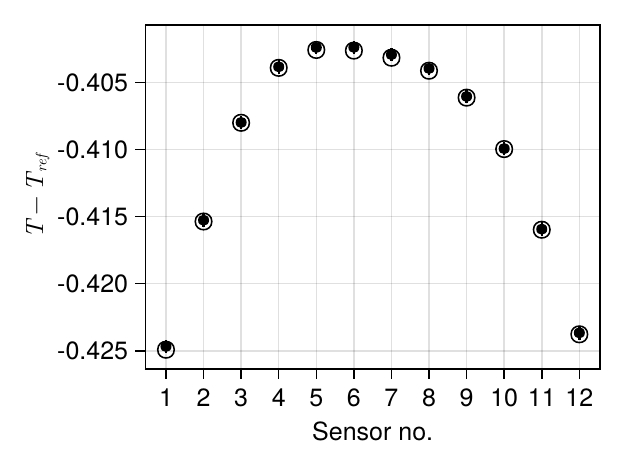} 
    \caption{}
    \label{fig:}
\end{subfigure}
\begin{subfigure}{0.3\textwidth}
    \centering
    \includegraphics[width=1\linewidth]{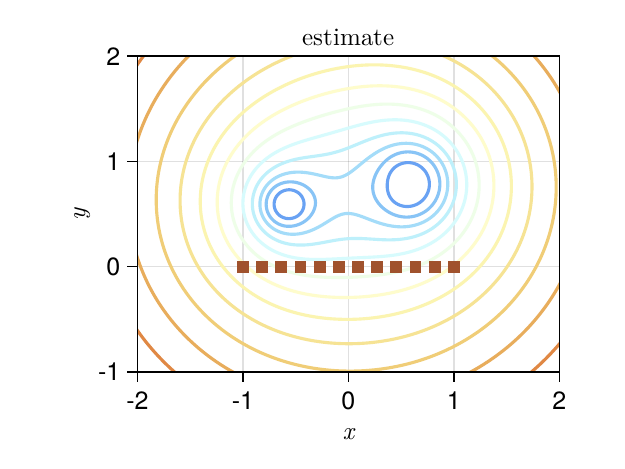} 
    \caption{}
    \label{fig:}
    
\end{subfigure}

    \caption{\centering Two true heaters and two heater estimators with twelve temperature measurements. The heaters' shapes are known, while their positions and strengths are inferred. (a) true temperature field, (b) the comparisons of sensor values between truth (filled circles) and estimate (open circles), (c) the estimated temperature field.}
    \label{fig:twoHeaters_12sensors}
\end{figure}

It is important to note that, in all scenarios, the inference reveals multiple possible solutions, each associated with a distinct peak likelihood, as illustrated in Figure \ref{fig:twoHeaters_sensors}(a). The mode with the maximum likelihood is considered superior among all possible solutions. Notably, the log-posterior values vary significantly, ranging from $-1.23$ to $-39.8$, indicating that the MCMC sampling may not have reached equilibrium. Employing the concept of maximum likelihood, we identify the mean of the Gaussian component with the highest likelihood, which is found to be $(\overline{x}_{o,1},\overline{y}_{o,1},\overline{q}_{h,1})=(0.5,0.79,1.0)$ and $(\overline{x}_{o,2},\overline{y}_{o,2},\overline{q}_{h,2})=(-0.61,0.59,1.97)$. The temperature field corresponding to the component with the highest likelihood is compared with the true temperature field in Figure \ref{fig:twoHeaters_12sensors}(a),(c), while Figure \ref{fig:twoHeaters_12sensors}(b) illustrates the true and estimated temperature distribution at the sensor locations. Notably, there is a robust correlation between the heat densities of the two heaters. The decrease in strength for one heater corresponds to an increase in strength for the other heater, ensuring a consistent temperature measurement at the sensor locations. The robust agreement between the true and estimated temperature results further attests to the reliability of the inference process.

\begin{figure}[htp!]
\begin{subfigure}{0.5\textwidth}
    \centering
    \includegraphics[width=1\linewidth]{figures/samples-2heater-1truth-K5-12sensor-ds018-noise5em4.pdf} 
    \caption{}
    \label{fig:}
\end{subfigure}
\begin{subfigure}{0.5\textwidth}
    \centering
    \includegraphics[width=1\linewidth]{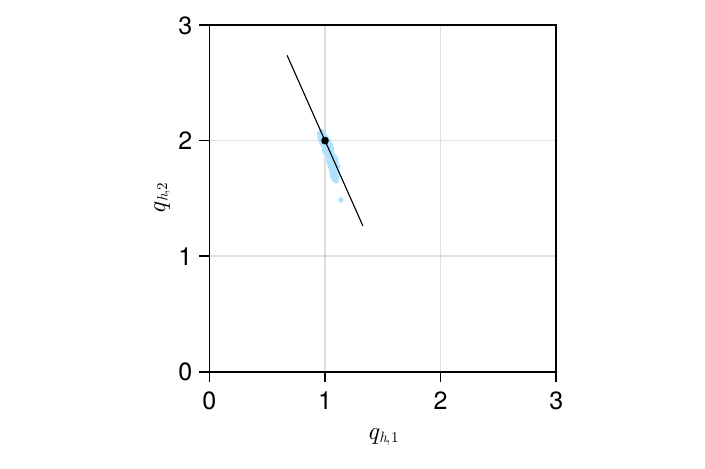} 
    \caption{}
    \label{fig:}
\end{subfigure}
\begin{subfigure}{0.5\textwidth}
    \centering
    \includegraphics[width=1\linewidth]{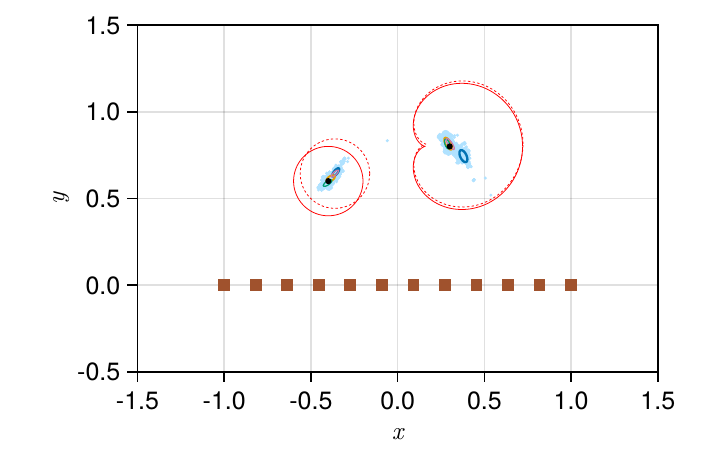} 
    \caption{}
    \label{fig:}
\end{subfigure}
\begin{subfigure}{0.5\textwidth}
    \centering
    \includegraphics[width=1\linewidth]{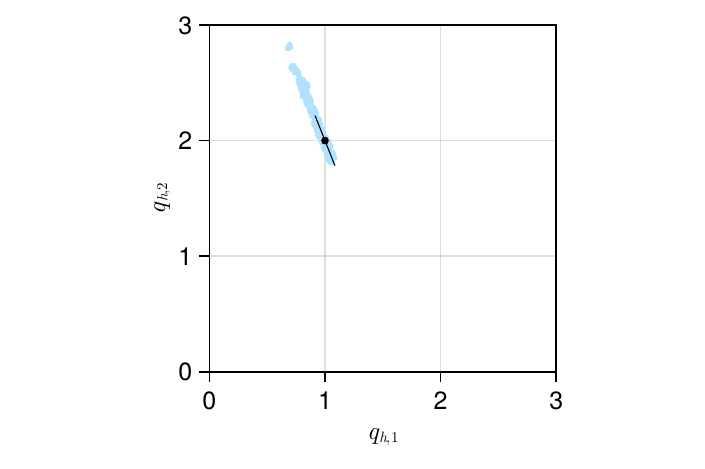} 
    \caption{}
    \label{fig:}
\end{subfigure}
\begin{subfigure}{0.5\textwidth}
    \centering
    \includegraphics[width=1\linewidth]{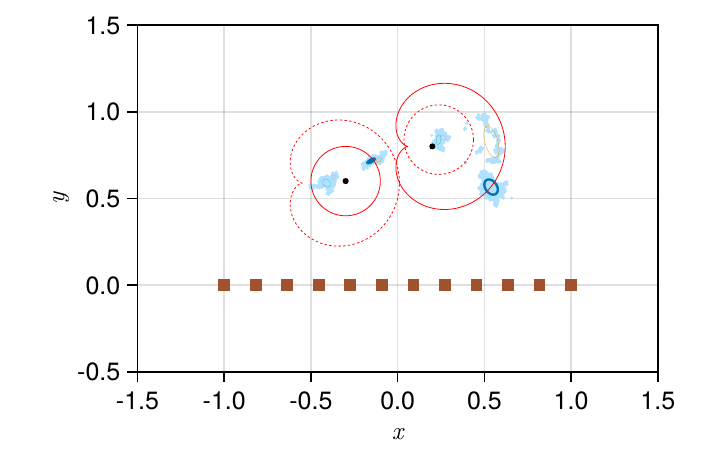} 
    \caption{}
    \label{fig:}
\end{subfigure}
\begin{subfigure}{0.5\textwidth}
    \centering
    \includegraphics[width=1\linewidth]{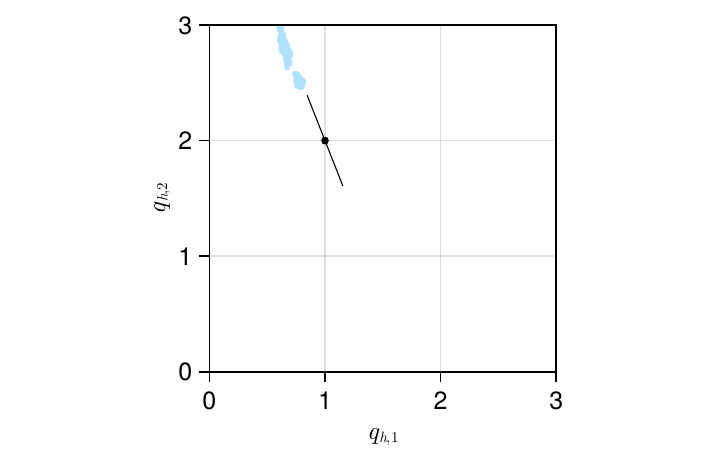} 
    \caption{}
    \label{fig:}
\end{subfigure}

    \caption{\centering Two true heaters and two heater estimators with twelve temperature measurements. The heaters' shapes are known, while their positions and strengths are inferred. The figures on the left panel represent the samples generated with MCMC, represented by five Gaussian components, for three different spacing between the two heaters. The true boundaries of the heaters are depicted by solid red curves, while the estimated heater modes with the greatest likelihood are illustrated by dashed curves. The right panel shows samples on $q_{h,1}-q_{h,2}$ corresponding to their left case. The black dots represent the true state and the black line illustrates the direction of maximum uncertainty.}
    \label{fig:twoHeaters_distance}
\end{figure}

The distance between the two heaters may influence the inference problem. To explore this, we bring the two heaters closer together in the $x$ direction while maintaining the average distance between them constant, $(x_{o,1}+x_{o,2})/2=-0.05$. Figure \ref{fig:twoHeaters_distance} displays the outcomes, where each line corresponds to a specific distance between the two heaters. The results indicate that, generally, as the heaters approach each other, there is an increase in uncertainty in estimating the strength of the heaters, as evident in the right panel of Figure \ref{fig:twoHeaters_distance}. Additionally, with closer heaters, multiple solutions emerge. For instance, Figure \ref{fig:twoHeaters_distance}(e) presents the true heaters alongside the best estimate with the highest likelihood, represented by solid and dashed red curves, respectively. In this case, the order of heaters has switched with displacement from the true centers' locations. The estimated heat densities $q_h$ also deviate from the truth, as depicted in Figure \ref{fig:twoHeaters_distance}(f). 

\subsection{Inference of a single heater with sensors placed on an adiabatic wall}
\begin{figure}[htp!]
    \centering
        \includegraphics[width=0.5\linewidth]{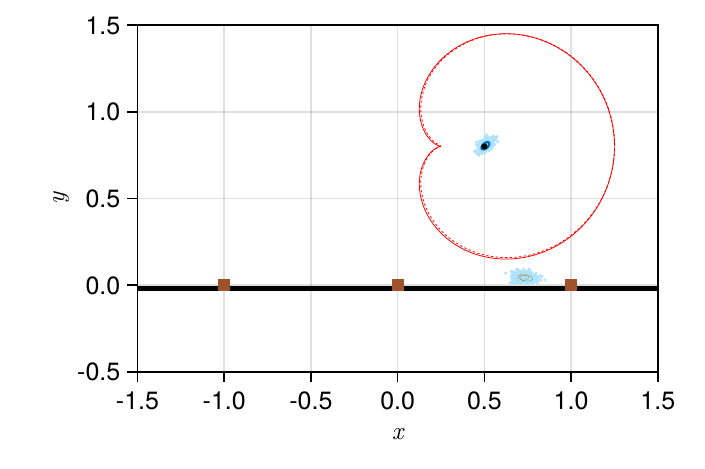} 
    \caption{\centering One true heater and one heater estimator with three temperature measurements, characterized by five Gaussian components for sensor noise equals to $5\times 10^{-3}$. Sensors are placed on a wall with an adiabatic boundary condition. The red curve illustrates the boundary of the heater.}
    \label{fig:samples_Neumann}
\end{figure}

Earlier sections addressed the inference problem of identifying one or multiple heat sources in an unbounded domain through temperature measurements. However, in practical scenarios, sensors are typically situated on a wall or a boundary. Consequently, a pertinent question remains unexplored: How do boundary conditions impact the performance of inference? If they do, is the influence positive or negative? To address these queries, we delve into the inverse problem of estimating the position and strength of a single heater centered at $(x_o,y_o) = (0.5,0.8)$ with unit strength and a known shape \emph{a priori} characterized by $(c_1,c_2) = (0.5,0.25)$ as the true states. Sensors are strategically positioned horizontally on an adiabatic wall, as illustrated in Figure \ref{fig:samples_Neumann} by a prominent black line beneath the square sensors. The measurement noise is set at $\sigma_\varepsilon=5 \times 10^{-3}$, akin to Figure \ref{fig:noise}(b). All conditions are similar to those of Figure \ref{fig:noise}(b) except that here, the sensors are placed on an adiabatic wall.

\begin{figure}[ht!]
\begin{subfigure}{0.3\textwidth}
    \centering
    \includegraphics[width=1\linewidth]{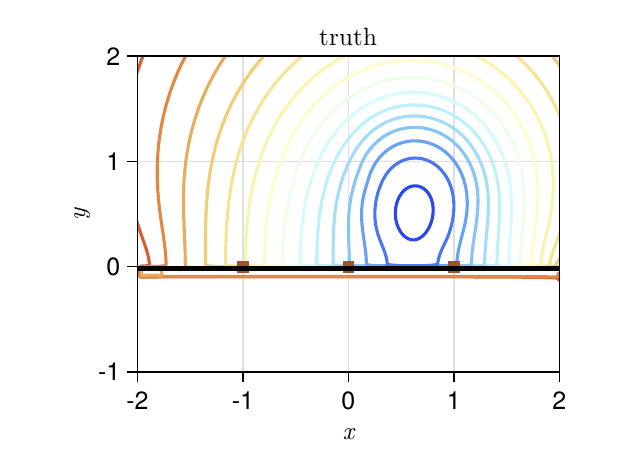} 
    \caption{}
    \label{fig:}
\end{subfigure}
\begin{subfigure}{0.3\textwidth}
    \centering
    \includegraphics[width=1\linewidth]{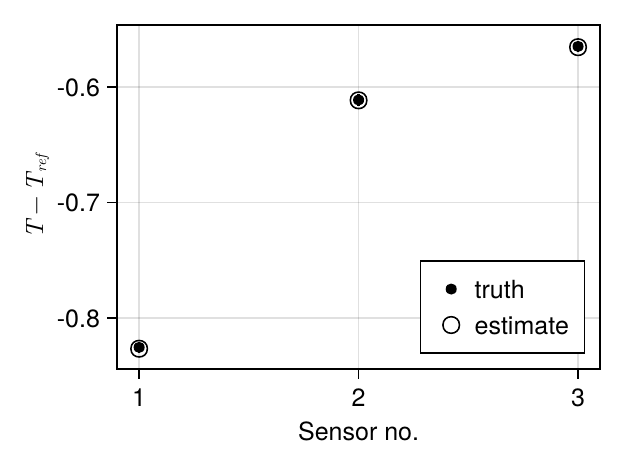} 
    \caption{}
    \label{fig:}
\end{subfigure}
\begin{subfigure}{0.3\textwidth}
    \centering
    \includegraphics[width=1\linewidth]{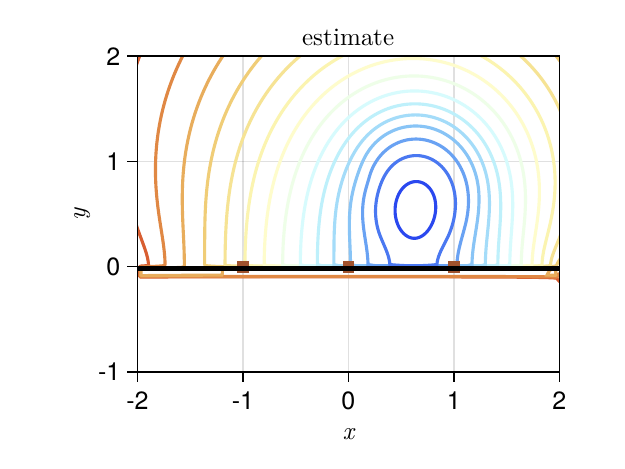} 
    \caption{}
    \label{fig:}
\end{subfigure}
\begin{subfigure}{1\textwidth}
    \centering
    \includegraphics[width=0.3\linewidth]{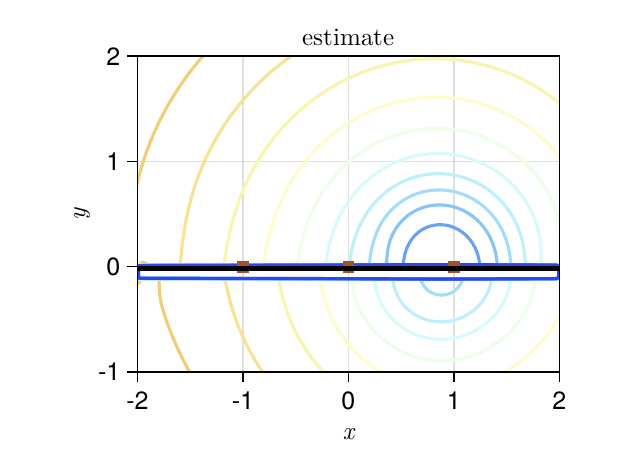} 
    \caption{}
    \label{fig:}
\end{subfigure}

    \caption{\centering One true heater and one heater estimator with 3 temperature measurements. The heater's shape is known, while its position and strength are inferred. (a) true temperature field, (b) the comparisons of sensor values between truth (filled circles) and estimate (open circles), and (c) the estimated temperature field. Panels (a)-(c) correspond to the estimated heater depicted in Figure \ref{fig:samples_Neumann}, while panel (d) is the temperature contour generated by the second solution.}
    \label{fig:Temp_NeumannBC}
\end{figure}

The results of the MCMC samples are visualized in Figure \ref{fig:samples_Neumann}. Two distinct solutions emerge: one closely aligns with the true state, while the other positions the heater close to the wall. In Figure \ref{fig:Temp_NeumannBC}, panels (a)-(c) showcase temperature contours and distributions at the sensor locations for the first solution corresponding to the estimated heater shown in Figure \ref{fig:samples_Neumann} with a dashed red curve, demonstrating excellent agreement with the ground truth. Conversely, the nature of the diffusion problem manifests in the second solution, where semi-circular temperature contours, nearly perpendicular to the wall, are observed (Figure \ref{fig:Temp_NeumannBC}(d)). Despite the smaller heat strength in the second solution, it yields identical temperature distributions at the sensor locations as the first solution. However, this second solution can be disregarded as it results in the heater intersecting the wall, which is physically implausible. Such spurious solutions can be easily eliminated by tailoring the bounding region in the prior.

Notably, the estimation for the unbounded domain exhibits greater uncertainty compared to the scenario where sensors are positioned along an adiabatic wall (compare the spread of MCMC samples in Figure \ref{fig:noise}(b) and Figure \ref{fig:samples_Neumann}). That can be attributed to higher temperatures at the location of sensors if they are placed on an adiabatic wall which results in lower noise-to-temperature ratios. In other words, the presence of the adiabatic wall can potentially enhance the estimation of the heater's characteristics.

\section{Conclusion} \label{conclusion}
This paper addresses the Bayesian inference of two-dimensional steady-state heat conduction in the presence of unknown heat sources. The goal is to infer the locations of the center of the heaters, along with their strengths, sizes and shapes, assimilating temperature observations in Euclidean space. The first three coefficients of the Fourier series in complex notation are considered in this paper to represent the heater's position and shape. The Markov Chain Monte Carlo (MCMC) method is employed to draw samples from the posterior distribution, and the random-walk Metropolis-Hasting (MH) algorithm is utilized to traverse the state space. To enhance exploration, the concept of parallel tempering is incorporated, allowing for potential exchange between chains. The inference problem is investigated in both an unbounded domain and a domain bounded at a wall with adiabatic boundary conditions where evaluation points are mounted.

A strong correlation exists between the heat source strength, $q_h$, and the area of the heater, $A$, mathematically expressed as $Q=q_hA$. This correlation underscores the behavior of the estimator in predicting infinite combinations of $q_h$ and $A$ that yield identical temperature observations at the evaluation points. This phenomenon highlights the potential for multiple solutions when estimating the total heat generation. Consequently, it is advisable to refrain from simultaneously estimating both the strength and the size of the heater. In our analysis, one of these variables is considered as \emph{a priori} known, with a sharply defined Gaussian distribution. Furthermore, it is observed that when the number of temperature sensors is less than the number of unknown states, the problem becomes rank deficient, leading to multiple solutions with high probability. By predefining the shape and dimensions of the heater as prior knowledge and deducing the location and strength of its center, adjusting the area while upholding a consistent total heat generation minimally affects the estimation of the heater's center. Nevertheless, when dealing with smaller-sized heaters, there is increased uncertainty surrounding $q_h$. This trend arises from the reduced heating moments associated with smaller heaters, contributing to the heightened uncertainty in estimating their strength.

The diffusive properties inherent in the steady-state heat conduction equation tend to smooth out temperature contours away from the heater's boundary. Consequently, when estimating the position and shape of the heater, this smoothing effect can lead to multiple solutions, causing ambiguity between a circular heater and the actual heater with a deformed shape. Furthermore, in the presence of multiple heaters, the convergence towards equilibrium slows down, and the emergence of multiple candidate solutions becomes more common, especially when the true heaters are positioned close to each other. 

When a heater is positioned within a domain bounded by a wall beneath the sensors, featuring Neumann boundary conditions for temperature, the estimation tends to achieve higher accuracy compared to scenarios where the heater is placed in an unbounded domain. This observation underscores the superior performance of estimation for heaters in confined domains, as they provide more information to the estimator. Investigating the impact of boundary locations on estimation performance presents an intriguing avenue for research. Boundaries situated far from the state-observation pair may exert lesser influence on the temperature distribution near the pair. A notable challenge in scenarios involving multiple heaters within the domain lies in the computational demands of Markov Chain Monte Carlo (MCMC) methods when exploring high-dimensional state spaces. This challenge can be mitigated by employing low-dimensional spaces through the estimation of fewer heaters. \cite{eldredge2023bayesian} successfully addressed this challenge in inferring point vortices from pressure measurements. When fewer heaters are involved in the estimation process, one might observe a tendency for like-sign heaters to aggregate in proximity to each other.

  \bibliographystyle{elsarticle-harv} 
  \bibliography{mybibliography}


%\begin{thebibliography}{00}



%\end{thebibliography}
\end{document}